\documentclass[runningheads]{llncs}
\usepackage{graphicx}
\usepackage{cvprabbreviation}
\usepackage{comment}
\usepackage{amsmath,amssymb} 
\usepackage{color}

\usepackage{booktabs}
\usepackage{subfloat}

\usepackage{float}
\usepackage{subfig}

\usepackage{bbding}
\usepackage{textcomp}
\usepackage{multirow}
\usepackage{overpic}
\usepackage{amsfonts,amssymb}
\usepackage{xcolor,colortbl}

\definecolor{Gray}{gray}{.90}

\newcommand{\tabincell}[2]{\begin{tabular}{@{}#1@{}}#2\end{tabular}}

\usepackage{algorithm}
\usepackage{algorithm}
\usepackage{algorithmicx}
\usepackage{algpseudocode}
\usepackage{amsmath}

\usepackage[pagebackref,breaklinks,colorlinks]{hyperref}

\usepackage[capitalize]{cleveref}
\crefname{section}{Sec.}{Secs.}
\Crefname{section}{Section}{Sections}
\Crefname{table}{Table}{Tables}
\crefname{table}{Tab.}{Tabs.}

\usepackage[accsupp]{axessibility}  

\begin{document}
\pagestyle{headings}
\mainmatter

\title{Self-Supervised Learning for Real-World Super-Resolution from Dual Zoomed Observations} 


\titlerunning{SelfDZSR}
%
%
\author{Zhilu Zhang\inst{1} \and
Ruohao Wang\inst{1} \and
Hongzhi Zhang\inst{1} $^{(}$\Envelope$^)$  \and
Yunjin Chen \and
Wangmeng Zuo\inst{1,2}}
\authorrunning{Z. Zhang et al.}
%
\institute{Harbin Institute of Technology, China \and Peng Cheng Laboratory \\
\email{\{cszlzhang, rhwangHIT\}@outlook.com, zhanghz0451@gmail.com, chenyunjin\_nudt@hotmail.com, wmzuo@hit.edu.cn}}
\maketitle

\begin{abstract}
  In this paper, we consider two challenging issues in reference-based super-resolution (RefSR), (i) how to choose a proper reference image, and (ii) how to learn real-world RefSR in a self-supervised manner. 
  Particularly, we present a novel self-supervised learning approach for real-world image SR from observations at dual camera zooms (SelfDZSR).
  Considering the popularity of multiple cameras in modern smartphones, the more zoomed (telephoto) image can be naturally leveraged as the reference to guide the SR of the lesser zoomed (short-focus) image. 
  Furthermore, SelfDZSR learns a deep network to obtain the SR result of short-focus image to have the same resolution as the telephoto image. 
  For this purpose, we take the telephoto image instead of an additional high-resolution image as the supervision information and select a center patch from it as the reference to super-resolve the corresponding short-focus image patch. 
  To mitigate the effect of the misalignment between short-focus low-resolution (LR) image and telephoto ground-truth (GT) image, we design an auxiliary-LR generator and map the GT to an auxiliary-LR while keeping the spatial position unchanged.
  Then the auxiliary-LR can be utilized to deform the LR features by the proposed adaptive spatial transformer networks (AdaSTN), and match the Ref features to GT.
  During testing, SelfDZSR can be directly deployed to super-solve the whole short-focus image with the reference of telephoto image.
  Experiments show that our method achieves better quantitative and qualitative performance against state-of-the-arts.
  Codes are available at \url{https://github.com/cszhilu1998/SelfDZSR}.
  %
  %
\end{abstract}

%
\section{Introduction}
\label{sec:intro}
  Image super-resolution (SR) aiming to recover a high-resolution (HR) image from its low-resolution (LR) counterpart is a severely ill-posed inverse problem with many practical applications. 
  To relax the ill-posedness of the SR, reference-based image SR (RefSR) has recently been suggested to super-resolve the LR image with more accurate details by leveraging a reference (Ref) image containing similar content and texture with the HR image.
  Albeit progress has been made in RefSR, it remains a challenging issue to choose a proper reference image for each LR image. 

  \begin{figure}[t]
	\centering
	\begin{overpic}
	  [width=0.95\linewidth]{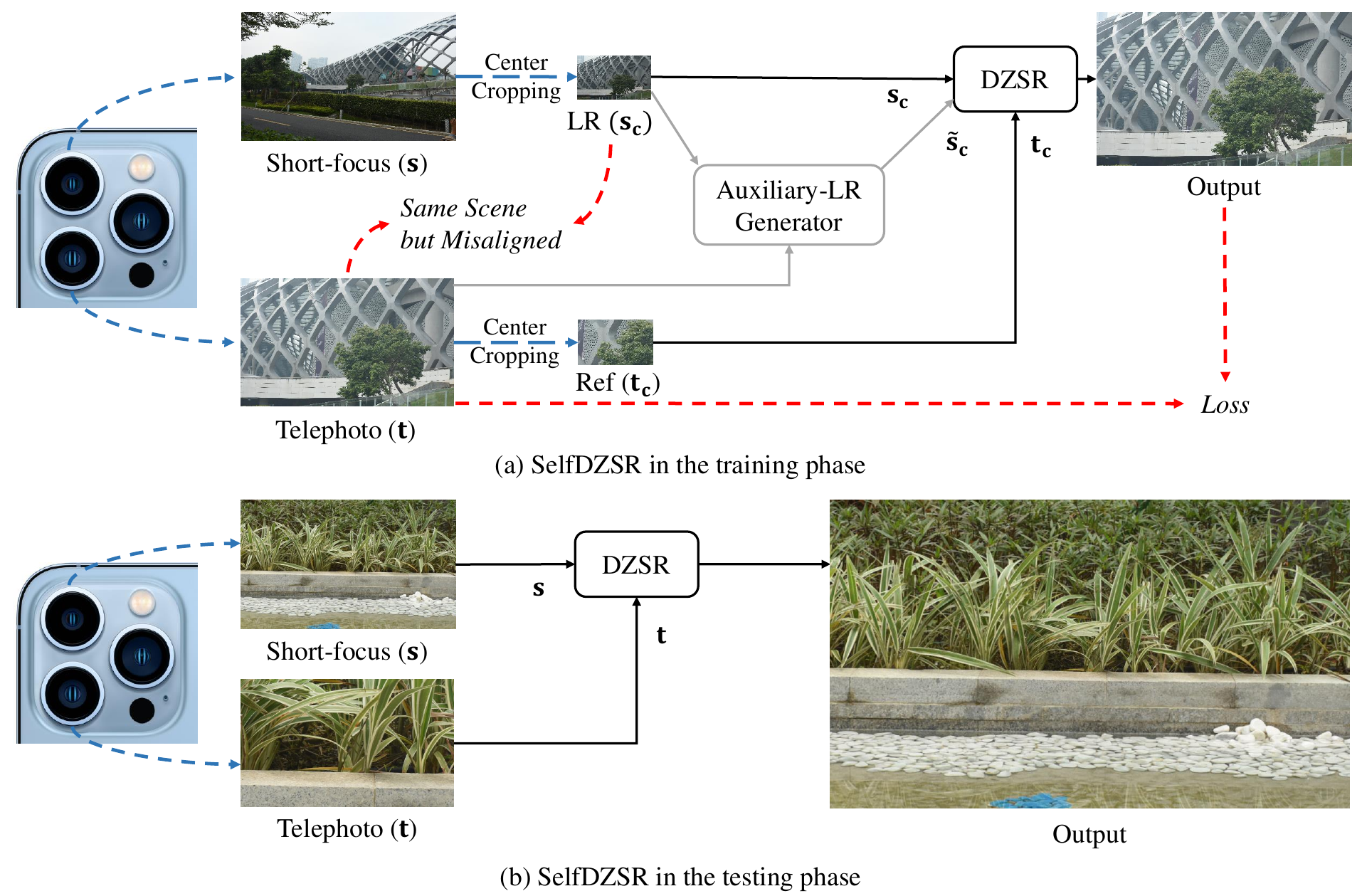}
	\end{overpic}
    \setlength{\abovecaptionskip}{1mm}
    \setlength{\belowcaptionskip}{-6mm}
	\caption{Overall pipeline of proposed SelfDZSR in the training and testing phase.}
	\label{fig:intro}
  \end{figure}
  Fortunately, advances and popularity of imaging techniques make it practically feasible to collect images of a scene at two different camera zooms (\ie, dual zoomed observations).
  For example, asymmetric cameras with different fixed-focal lenses have been equipped in modern smartphones.
  In these practical scenarios, the more zoomed (telephoto) image can be naturally leveraged as the reference to guide the SR of the lesser zoomed (short-focus) image. 
  Image SR from dual zoomed observations (DZSR) can thus be regarded as a special case of RefSR, in which Ref has the same scene as the center part of the LR image but is with higher resolution. 
  While conventional RefSR methods~\cite{SRNTT,FRM,SSEN,TTSR,MASA-SR,C2-Matching} usually use synthetic (\eg, bicubic) degraded LR images for training and evaluation, DZSR should cope with real-world LR short-focus images and no ground-truth HR images are available in training. 
  To bridge the domain gap between synthetic and real-world LR images, dual-camera super-resolution (DCSR)~\cite{DCSR} suggests self-supervised real-image adaptation (SRA) involving a degradation preserving and a detail transfer terms. 
  However, DCSR only attains limited success, due to that the two loss terms in SRA cannot well address the gap between the synthetic and real-world LR degradation as well as the misalignment between short-focus and telephoto images.

  In this work, we aim at super-resolving the real world short-focus image with the reference of the corresponding telephoto image.
  Different from DCSR~\cite{DCSR} requiring to pre-train on synthetic images, we adopt self-supervised learning to train DZSR model (\ie, SelfDZSR) from scratch directly on short-focus and telephoto images without additional high-resolution images.
  As shown in Fig.~\ref{fig:intro}(a), instead of the whole images, during training we crop the center part of the short-focus and telephoto image respectively as the input LR and Ref, and use the whole telephoto image as the ground-truth (GT). 
  In the testing phase, by using the whole short-focus and telephoto images respectively as LR and Ref, SelfDZSR can be directly deployed to super-solve the whole short-focus image, as shown in Fig.~\ref{fig:intro}(b).
  
  However, when training SelfDZSR, the cropped short-focus image generally cannot be accurately aligned with the telephoto GT image, making the learned model prone to producing blurry SR results~\cite{RAW-to-sRGB,SRRAW}.  
  Matching the Ref to LR will also result in the warped Ref being not aligned with the GT, bringing more uncertainty to network training.
  To handle the misalignment issue, we map GT to an auxiliary-LR while keeping the spatial position unchanged, and utilize the auxiliary-LR as the target position image for deforming LR and Ref features during training.
  During testing, given that GT is unavailable, the auxiliary-LR should be replaced by LR.
  Specifically, we propose an auxiliary-LR generator network constrained by position preserving loss and content preserving loss. 
  The position preserving loss constrains that the auxiliary-LR is aligned with GT, so that the warped LR and Ref features are aligned with GT during training. 
  The content preserving loss constrains that the auxiliary-LR has similar contents as LR, so that the auxiliary-LR can be replaced by LR safely during testing.
  
  Moreover, for aligning LR with GT, we propose adaptive spatial transformer networks (AdaSTN) that takes the auxiliary-LR and LR images for estimating the offsets between them to deform the LR features.
  AdaSTN implicitly aligns contents between LR and auxiliary-LR by minimizing the reconstruction loss of SelfDZSR.
  When training is done, the auxiliary-LR generator and the offset estimator of AdaSTN can be safely detached.
  That is, the estimated offsets of AdaSTN can be set to default, bringing no extra cost in the test phase.
  For the matching of Ref image, instead of searching corresponding contents from Ref to LR features in most existing RefSR methods~\cite{SRNTT,TTSR,MASA-SR,C2-Matching,DCSR}, we perform it from Ref to auxiliary-LR features.
  Finally, the warped LR and warped Ref features can be regarded as aligned with GT, which are then combined and fed into the restoration module.

  Extensive experiments are conducted on the Nikon camera from the DRealSR dataset~\cite{CDC} as well as the CameraFusion dataset~\cite{DCSR}. 
  The results demonstrate the effectiveness and practicability of our SelfDZSR for real-world image SR from dual zoomed observations.
  In comparison to the state-of-the-art SR and RefSR methods, our SelfDZSR performs favorably in terms of both quantitative metrics and perceptual quality.

  To sum up, the main contributions of this work include:
  \begin{itemize}
    \item An effective self-supervised learning approach, \ie, SelfDZSR, is presented to super-resolve the real-world images from dual zoomed observations.
    \item The adverse effect of image misalignment for self-supervised learning and Ref matching is alleviated by the proposed auxiliary-LR and adaptive spatial transformer networks (AdaSTN), while bringing no extra inference cost. %
    \item Quantitative and qualitative results on Nikon camera and CameraFusion show that our method outperforms the state-of-the-art methods. 
  \end{itemize}
%
\section{Related Work} 

\subsection{Blind Image Super-Resolution}
  %
  With the development of deep networks, single image super-resolution (SISR) methods based on fixed and known degradation have achieved great success in terms of both performance~\cite{SRCNN,SRGAN,EDSR,RCAN,SRMD} and efficiency~\cite{IMDN,AdaDSR,wang2021exploring,ClassSR,Xie_2021_ICCV}.
  However, these methods perform poorly when applying to images with an unknown degradation, and may cause some artifacts.
  Thus, blind super-resolution comes into being to bridge the gap.
  
  On the one hand, some works estimate the blur kernel or degradation representation for LR and feed it into the SR reconstruction network.
  IKC~\cite{IKC} performed kernel estimation and SR reconstruction processes iteratively, while DAN~\cite{DAN} conducted it in an alternating optimization scheme.
  KernelGAN~\cite{KerGAN} utilized the image patch recurrence property to estimate an image-specific kernel, and FKP~\cite{FKP} learned a kernel prior based on normalization flow~\cite{flow} at test time.
  To relaxing the assumption that blur kernels are spatially invariant, MANet~\cite{MANet} estimated spatially variant kernel by suggesting mutual affine convolution.
  Different from the above explicit methods of estimating kernel, DASR~\cite{DASR} introduced contrastive learning~\cite{he2020momentum} to extract discriminative representations to distinguish different degradations. 
  On the other hand, Hussein~\etal~\cite{correction_cvpr2020} modified the LR to a pre-defined degradation (\eg, bicubic) type by a closed-form correction filter.
  BSRGAN~\cite{BSRGAN} and Real-ESRGAN~\cite{Real-ESRGAN} designed more complex degradation models to generate LR data for training the networks, making the networks generalize well to many real-world degradation scenarios.
  %

\subsection{Real-World Image Super-Resolution} 
  Although blind SR models trained on synthetic data have shown appreciable generalization capacity, the formulated degradation assumption limits the performance on real-world images with much more complicated and changeable degradation.
  Thus, image SR directly towards real-world scenes has also received much attention.
  On the one hand, given unpaired real LR an HR, several real-world SR methods~\cite{NTIRE2019,AIM2020,wei2021unsupervised} attempt to approximate real degradation and generate the auxiliary-LR image from HR, and then learn to super-resolve the auxiliary-LR in a supervised manner.
  On the other hand, some methods~\cite{CameraLen,LP-KPN,CDC,AIM2019,NTIRE2020} construct paired datasets by adjusting the focal length of a camera, in which the image with a long focal and short focal length is regarded as GT and LR, respectively.
  Among these methods, LP-KPN~\cite{LP-KPN} presented a kernel prediction network based on the Laplacian pyramid.
  CDC~\cite{CDC} considered reconstruction difficulty of different components, and performed image SR in a divide-and-conquer manner.

  In addition, the misalignment of data pairs is a universal problem in real scenes, and it may cause blurry SR result.
  The above methods based on paired datasets pre-execute complex alignment or even manual selection, which are generally laborious and time-consuming.
  Different from them, CoBi~\cite{SRRAW} loss offered an effective way to deal with misalignment during SR training.
  Zhang~\etal~\cite{RAW-to-sRGB} incorporated global color mapping and optical flow~\cite{PWC-Net} to explicitly align the data pairs with severe color inconsistency.
  Nevertheless, optical flow is limited in handling complicated misalignment.
  In this work, we further propose AdaSTN to handle the complicated misalignment after pre-alignment with optical flow.

\subsection{Reference-Based Image Super-Resolution}
  RefSR aims to take advantage of a high-resolution reference image that has similar content and texture as LR for super-resolution.
  It relaxes the ill-posedness of SISR and facilitates the generation of more accurate details.
  The features extracting and matching between LR and Ref is the research focus of most RefSR methods.
  Among them, Zheng~\etal~\cite{refsr_bwvc} proposed a correspondence network to extract features for matching, and an HR synthesis network with the input of the matched Ref.
  SRNTT~\cite{SRNTT} calculated the correlation between pre-trained VGG features of LR and Ref at multiple levels for matching them.
  Zhang~\etal~\cite{zhang2020texture} extended the scaling factor of RefSR methods from $4\times$ to $16\times$.
  Furthermore, TTSR~\cite{TTSR} and FRM~\cite{FRM} developed an end-to-end training framework and proposed learnable feature extractors.
  Recently, $C^2$-Matching~\cite{C2-Matching} performed a more accurate match by the teacher-student correlation distillation while MASA-SR~\cite{MASA-SR} reduced the computational cost by coarse-to-fine correspondence matching.
  Besides, CrossNet~\cite{CrossNet} and SEN~\cite{SSEN} respectively introduced optical flow~\cite{FlowNet} and deformable convolution~\cite{DefConv,DefConv-v2} to align Ref with LR.
  However, optical flow is limited in handling large and complicated motions while deformable convolution is limited in modeling long-distance correspondence.
  In this work, we follow~\cite{C2-Matching} to perform patch-wise matching.

  Additionally, the RefSR methods mentioned above are all based on the bicubic down-sampling.
  DCSR~\cite{DCSR} explores an adaptive fine-tuning strategy on real-world images based on the pre-trained model with synthetic data.
  In this work, we propose a fully self-supervised learning framework directly on weakly aligned dual zoomed observations. 
  
\section{Proposed Method} 
  In this section, we first give a description of our self-supervised learning framework for super-resolution from dual zoomed observations. 
  Then we detail the solutions for handling the misalignment problem, including the generation of auxiliary-LR, alignment between LR and auxiliary-LR by AdaSTN, alignment between Ref and (auxiliary-)LR.
  Finally, the design of the restoration module is introduced, and the learning objective is provided.

\subsection{Self-Supervised Learning Framework} \label{self-supervised}
  %
  Denote by $\mathbf{s}$ and $\mathbf{t}$ the short-focus image and the telephoto image, respectively.
  Super-resolution based on dual zoomed observations aims to super-resolve the short-focus image $\mathbf{s}$ with the reference telephoto image $\mathbf{t}$, which can be written as, 
    \begin{equation}
      \setlength{\abovedisplayskip}{1.5mm}
      \setlength{\belowdisplayskip}{1.5mm}
      \hat{\mathbf{y}} = \mathcal{Z}(\mathbf{s}, \mathbf{t}; \Theta_\mathcal{Z}),
    \label{eqn:DZSR}
    \end{equation}
  where $\hat{\mathbf{y}}$ has the same field-of-view as $\mathbf{s}$ and the same resolution as $\mathbf{t}$, $\mathcal{Z}$ denotes the zooming network with the parameter $\Theta_\mathcal{Z}$.
  
  However, in real-world scenarios, the GT of $\hat{\mathbf{y}}$ is hard or almost impossible to acquire.
  A simple alternative solution is to leverage synthetic data for training, but the domain gaps between the degradation model in training and that in real-world images prevent it from working well.
  DCSR~\cite{DCSR} tries to bridge the gaps by fine-tuning the trained model using an SRA strategy, but the huge difference in the field of view between the output and the target telephoto image limits it in achieving satisfying results.

  In contrast to the above methods, we propose a novel self-supervised dual-zooms super-resolution (SelfDZSR) framework, which can be trained from scratch solely on the short-focus and telephoto image (see Fig.~\ref{fig:intro}(a)), and be directly deployed to the real-world dual zoomed observations (see Fig.~\ref{fig:intro}(b)).
  During training, we first crop the central area of the telephoto and short-focus images,
    \begin{equation}
      \setlength{\abovedisplayskip}{1.5mm}
      \setlength{\belowdisplayskip}{1.5mm}
      \mathbf{s_c} = \mathcal{C}(\mathbf{s}; r), \qquad
      \mathbf{t_c} = \mathcal{C}(\mathbf{t}; r),
    \label{eqn:crop}
    \end{equation}
  where $\mathcal{C}$ denotes the center cropping operator, $r$ is the focal length ratio of $\mathbf{t}$ and $\mathbf{s}$.
  Note that $\mathbf{t_c}$ has the same scene and higher resolution with $\mathcal{C}(\mathbf{s_c}; r)$, \ie, the central area of $\mathbf{s_c}$.
  Moreover, $\mathbf{t}$ and $\mathbf{s_c}$ have the same scene, while the resolution is different by $r$ times.
  Thus, $\mathbf{s_c}$ and $\mathbf{t_c}$ can be naturally used as LR and Ref respectively, while $\mathbf{t}$ can be regarded as the GT.
  Then we can define DZSR as,
    \begin{equation}
      \setlength{\abovedisplayskip}{1.5mm}
      \setlength{\belowdisplayskip}{1.5mm}
      \Theta_\mathcal{Z} = \arg \min_{\Theta_\mathcal{Z}} \mathcal{L}\left(\mathcal{Z}(\mathbf{s_c}, \mathbf{t_c}; \Theta_\mathcal{Z}), \mathbf{t} \right)
    \label{eqn:SelfDZSR}
    \end{equation}
  where $\mathcal{L}$ denotes the self-supervised learning objective.

  Nonetheless, GT $\mathbf{t}$ is not spatially aligned with LR $\mathbf{s_c}$, bringing adverse effects on self-supervised learning and Ref $\mathbf{t_c}$ matching.
  To handle the misalignment issue, we hope to construct an auxiliary-LR to guide the deformation of LR and Ref towards the GT.
  For this purpose, the elaborate design on network architecture and loss terms is essential for SelfDZSR, which is introduced below. 

   \begin{figure}[t]
	\centering
	\begin{overpic}
	  [width=0.98\linewidth]{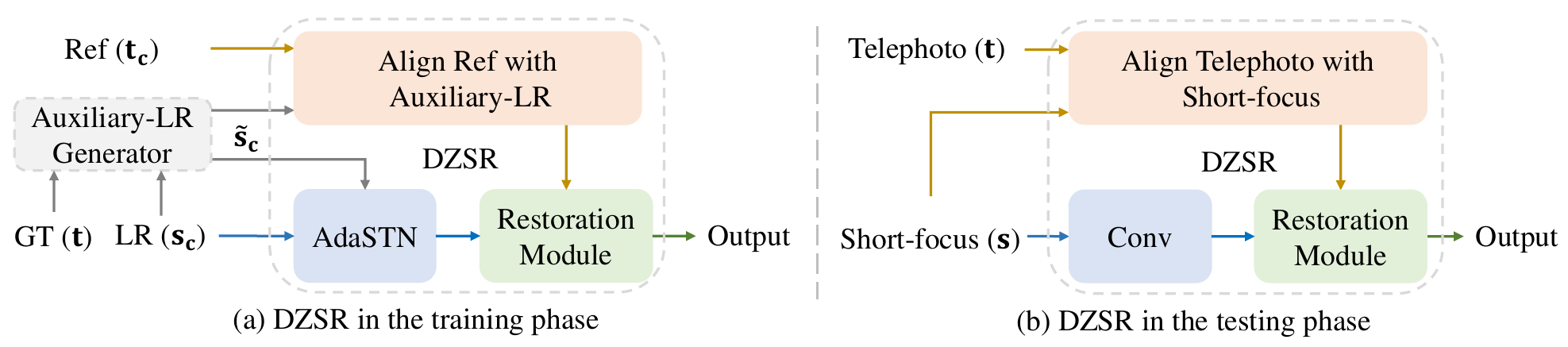}
	\end{overpic}
    \setlength{\abovecaptionskip}{0.5mm}
    \setlength{\belowcaptionskip}{-2mm}	
    \caption{The pipeline of proposed DZSR model. 
	(a)~DZSR in the training phase. The auxiliary-LR is generated to guide the deformation of LR and Ref towards the GT. The aligned LR and Ref features are fed into the restoration module. 
	(b)~DZSR in the testing phase. The short-focus and telephoto image can be regarded as LR and Ref, respectively.
    The auxiliary-LR generator is detached and AdaSTN is simplified to a convolution layer. } 
	\label{fig:SelfDZSR}
  \end{figure}
  
\subsection{Generation of Auxiliary-LR for Alignment} \label{sec:auxiliary-LR_gen}
  %
  For SelfDZSR during training, the LR image $\mathbf{s_c}$ and GT image $\mathbf{t}$ are captured from the different camera lenses, and thus are misaligned in space.  
  It has been shown in recent works~\cite{SRRAW,RAW-to-sRGB} that the spatial misalignment of data pairs will cause the network to produce blurry results.
  Off-the-shelf optical flow~\cite{PWC-Net} offers a probable solution in dealing with this issue.
  However, limited to the offset diversity~\cite{chan2021understanding} of optical flow, images after registration are still slightly misaligned in some complex circumstances and explicit perfect alignment is very difficult. 
  Moreover, the misalignment will result in the warped Ref features being not aligned with GT after matching Ref to LR, bringing more uncertainty to network training. 
  To handle the above issues, we construct an auxiliary-LR $\mathbf{\tilde{s}_c}$ from the GT $\mathbf{t}$ while keeping the spatial position unchanged, and take it to guide the alignment of LR and Ref towards GT (see Fig.~\ref{fig:SelfDZSR}(a)).
  Noted that the auxiliary-LR cannot be used in testing, and it should be substituted by the short-focus $\mathbf{s}$ (see Fig.~\ref{fig:SelfDZSR}(b)).
  
  \begin{figure}[t]
    \centering
    \begin{overpic} 
      [width=0.98\linewidth]{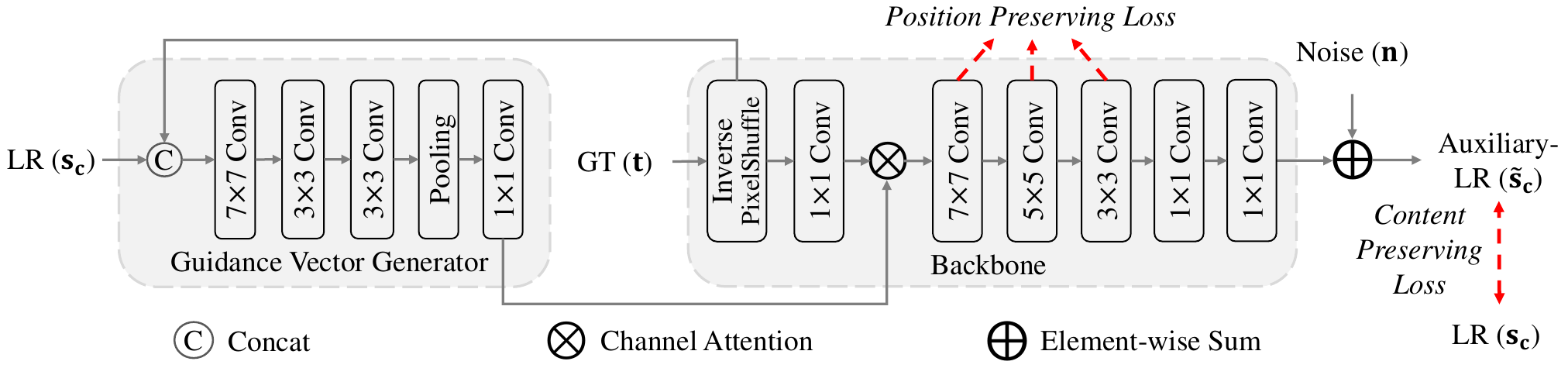}
    \end{overpic}
    \setlength{\abovecaptionskip}{0mm}
    \setlength{\belowcaptionskip}{-6mm}	
    \caption{Illustration of the auxiliary-LR generator. The position preserving loss constraints the kernel weight to ensure the alignment between auxiliary-LR and GT, while content preserving loss constraints that auxiliary-LR has similar contents as LR.}
    \label{fig:degrade}
  \end{figure}
  
  Thus, the auxiliary-LR $\mathbf{\tilde{s}_c}$ is required to satisfy two prerequisites.
  (i) $\mathbf{\tilde{s}_c}$ can be substituted by $\mathbf{s}$ during testing.
  (ii) The spatial position of $\mathbf{\tilde{s}_c}$ should keep the same as $\mathbf{t}$.
  For the first point, The auxiliary-LR should have similar contents and degradation types as LR, so that it can be substituted safely during testing.
  In particular, we design an auxiliary-LR generator network and constrain the contents of auxiliary-LR to be similar with these of LR, as shown in Fig.~\ref{fig:degrade}.
  For the second point, inspired by KernelGAN~\cite{KerGAN}, we take advantage of the position preserving loss to constrain the centroid of the local convolution kernel in the center of space. 
  The position preserving loss $\mathcal{L}_\mathrm{p}$ can be defined as,
  \begin{equation}
      \setlength{\abovedisplayskip}{1.5mm}
      \setlength{\belowdisplayskip}{1.5mm}
    \begin{split}
    \mathcal{L}_\mathrm{p}(\mathbf{W^\mathit{l}}) &  =  \| \sum_{i=0}^{k-1}\sum_{j=0}^{k-1} (i- \frac{k}{2} + 0.5) \mathit{w_{i,j}^{l}} \|_1  + \| \sum_{i=0}^{k-1}\sum_{j=0}^{k-1} (j - \frac{k}{2} + 0.5)  \mathit{w_{i,j}^{l}} \|_1 ,
    \end{split}
    \label{eqn:loss_center}
  \end{equation}  
  where $\mathbf{W^\mathit{l}}$ denotes the kernel weight parameters of the $\mathit{l}$-th convolution layer in the backbone of the auxiliary-LR generator, $k$ is odd and denotes the kernel size, $\mathit{w_{i,j}^{l}}$ denotes the value in the $\mathit(i,j)$ position of $\mathbf{W^\mathit{l}}$.
  In addition, LR $\mathbf{s_c}$ can be used to generate a conditional guidance vector for modulating features of $\mathbf{t}$ globally.
  Noted that the global modulation does not affect the preservation of spatial position.
  In short, the auxiliary-LR can be represented as,
    \begin{equation}
    \setlength{\abovedisplayskip}{1.5mm}
    \setlength{\belowdisplayskip}{1.5mm}
      \mathbf{\tilde{s}_c} = \mathcal{D}(\mathbf{t}, \mathbf{s_c}; \Theta_\mathcal{D}) + \mathbf{n},
    \label{eqn:degrad}
    \end{equation}
  where $\mathcal{D}$ denotes the auxiliary-LR generator network with the parameter $\Theta_\mathcal{D}$, $\mathbf{n}$ denotes the noise detailed in the suppl.
  Furthermore, $\Theta_\mathcal{D}$ can be written as,
  \begin{equation}
    \setlength{\abovedisplayskip}{1.5mm}
    \setlength{\belowdisplayskip}{1.5mm}
    \Theta_\mathcal{D} = \arg \min_{\Theta_\mathcal{D}} \|(\mathbf{\tilde{s}_c} - \mathbf{n}) - \mathbf{s_c}\|_1  + \lambda_\mathit{p}\sum_{l=1}^L\mathcal{L}_\mathrm{p}(\mathbf{W^\mathit{l}}),
    \label{eqn:loss_d}
  \end{equation}
  where we set $\lambda_\mathit{p}$ to 100.

\subsection{Alignment between LR and GT by AdaSTN} \label{sec:alignment_gt}
  \begin{figure}[t]
    \centering
    \begin{overpic} 
      [width=0.8\linewidth]{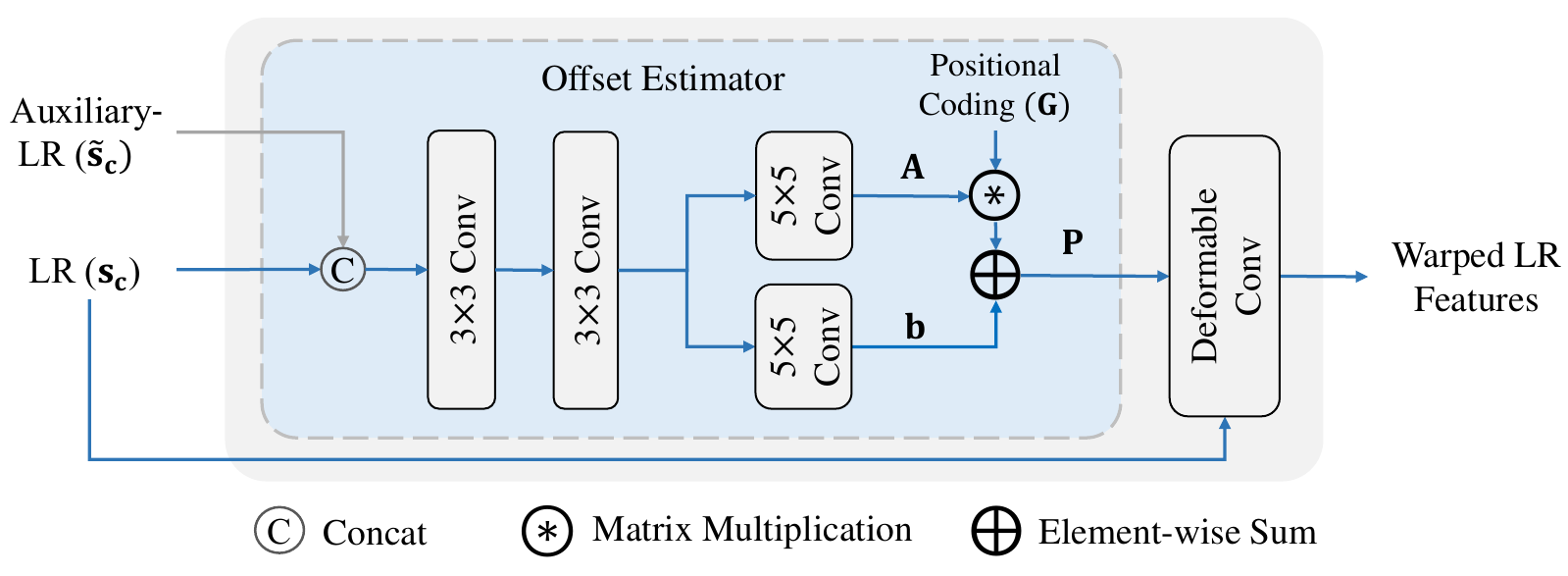}
    \end{overpic}
    \setlength{\abovecaptionskip}{0mm}
    \setlength{\belowcaptionskip}{-6mm}	
    \caption{Illustration of AdaSTN. } 
    \label{fig:AdaSTN}
  \end{figure}
  Given LR and auxiliary-LR, we suggest implicitly aligning LR to auxiliary-LR (aligned with GT).
  We can estimate the offsets between them and then deform the LR features to align with GT.
  Deformable convolution~\cite{DefConv} is a natural choice, but the direct estimation of the offsets may bring instability to the network training.
  Inspired by~\cite{AdaConv}, we propose adaptive spatial transformer networks (AdaSTN) that offset is obtained indirectly by estimating the pixel-level affine transformation matrix and translation vector, as shown in Fig.~\ref{fig:AdaSTN}.
  For every pixel, the estimated offset of AdaSTN can be written as,
    \begin{equation}
      \setlength{\abovedisplayskip}{1.5mm}
      \setlength{\belowdisplayskip}{1.5mm}
      \mathbf{P} = \mathbf{A}\mathbf{G} + \mathbf{b}, 
    \label{eqn:offset}
    \end{equation}
  where $\mathbf{A}\in\mathbb{R}^{2\times2}$ is a predicted affine transformation matrix and $\mathbf{b}\in\mathbb{R}^{2\times1}$ is the translation vector.
  $\mathbf{G}$ is a positional coding represented by
    \begin{equation}
      \setlength{\abovedisplayskip}{1.5mm}
      \setlength{\belowdisplayskip}{1.5mm}
      \mathbf{G} = \begin{bmatrix}
                   -1 & -1 & -1 & 0  & 0 & 0 & 1  & 1 & 1 \\
                   -1 & 0  & 1  & -1 & 0 & 1 & -1 & 0 & 1
                   \end{bmatrix}.
    \label{eqn:G}
    \end{equation}
  Thus, the deformable convolution of AdaSTN can be formulated as,
    \begin{equation}
      \setlength{\abovedisplayskip}{1.5mm}
      \setlength{\belowdisplayskip}{1.5mm}    
       \mathbf{y}(\mathbf{q}) = \sum\nolimits_{k=0}^8\mathbf{w}_k \mathbf{x}(\mathbf{q} + \mathbf{p}_k), 
    \label{eqn:defconv}
    \end{equation}
    where $\mathbf{x}$ and $\mathbf{y}$ represent the input and output features, respectively.
    $\mathbf{w}_k$ denotes the kernel weight and $\mathbf{p}_k$ denotes the $k$-th column value of $\mathbf{P}$.
    AdaSTN can be regarded as a variant of STN~\cite{STN}, which is from a global mode to a pixel-wise mode.
    In comparison to deformable convolution~\cite{DefConv}, AdaSTN is more stable in estimating the offsets.
    During experiments, we stack 3 AdaSTNs to align LR and auxiliary-LR progressively.
    
    Note that auxiliary-LR is not available in the testing phase.
    A feasible way is to replace auxiliary-LR with LR.
    Actually, there is no need to estimate the offsets for AdaSTN.
    We can set $\mathbf{P} = \mathbf{0}$ directly, which means that the deformable convolution of AdaSTN can only observe the input value at the center point of the kernel and AdaSTN degenerates into 1$\times1$ convolution (see Fig.~\ref{fig:SelfDZSR}(b)).
    However, this way may produce some artifacts in the results due to the gap between training and testing.
    In order to bridge this gap, for each AdaSTN, we randomly set $\mathbf{P} = \mathbf{0}$ with probability $p$ (\eg, 0.3) during training. 
    For each training sample, the probability $p^3$ (\eg, 0.027) that 3 AdaSTNs are all set to $\mathbf{P} = \mathbf{0}$ is low, so it has little impact on the learning of the overall framework.
    %
  
\subsection{Alignment between Ref and (Auxiliary-)LR} \label{sec:alignment_ref}
\begin{figure}[t]
    \centering
    \begin{overpic} 
      [width=.85\linewidth]{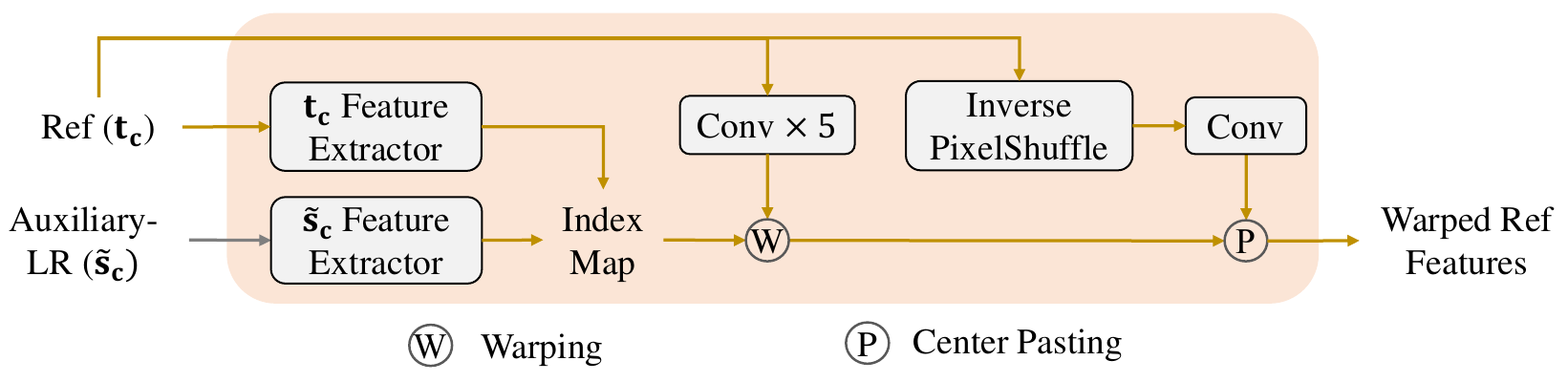}
    \end{overpic}
    \setlength{\abovecaptionskip}{0mm}
    \setlength{\belowcaptionskip}{-6mm}	
    \caption{Alignment between Ref and auxiliary-LR.}
    \label{fig:alignref}
  \end{figure}

  %
  %
  Previous RefSR methods generally perform matching by calculating cosine similarity between Ref and LR features.
  For SelfDZSR during training, the misalignment between LR and GT will result in the warped Ref features being not aligned with GT after matching Ref to LR.
  %
  To handle the issue, we instead calculate the correlation between Ref and auxiliary-LR features (see Fig.~\ref{fig:SelfDZSR}(a)).
  And the auxiliary-LR $\mathbf{\tilde{s}_c}$ can be substituted by the short-focus $\mathbf{s}$ during testing (see Fig.~\ref{fig:SelfDZSR}(b)).
  Fig.~\ref{fig:alignref} shows the alignment between Ref and auxiliary-LR.
  The index map is obtained by calculating the cosine similarity between Ref and auxiliary-LR features that are extracted by pre-trained feature extractors.
  Then the Ref is warped according to the index map.
  In addition, for SelfDZSR, the central part of LR has the same scene as Ref.
  Taking this property into account, we can rearrange Ref elements by an inverse PixelShuffle~\cite{PixelShuffle} layer, and then paste it to the center area of the warped Ref features.
  %
  %

\subsection{Restoration Module and Learning Objective}  \label{sec:restoration}
  %
  \noindent{}\textbf{Restoration Module}.
  After getting the aligned LR and aligned Ref features, we concatenate and feed them into the backbone of the restoration module consisting of 16 residual blocks~\cite{EDSR}.
  The concatenated features are also fed into an encoder to generate condition vectors for global modulation of the backbone features.
  The role of modulation is to make better use of Ref information and relieve the color inconsistency between LR and GT images.
  The details of restoration module will be introduced in the suppl.
  
  \noindent{}\textbf{Learning Objective for DZSR}.
  The sliced Wasserstein (SW) distance has exhibited  outstanding merit for training deep generative networks~\cite{SWDGen1,SWDGen2}.
  Recently, SW loss has been successfully applied in texture synthesis~\cite{SWLoss1}, image enhancement~\cite{SWLoss2} and \etc.
  Here, we also use SW loss $\mathcal{L}_\mathrm{SW}$ to optimize DZSR, the detailed description will be given in the suppl.
  The loss term of DZSR can be written as,
  \begin{equation}
      \setlength{\abovedisplayskip}{1.5mm}
      \setlength{\belowdisplayskip}{1.5mm}  
        \mathcal{L}_\mathrm{DZSR}(\hat{\mathbf{y}}, \mathbf{t})
        = \|\hat{\mathbf{y}}-\mathbf{t}\|_1 
        + \lambda_\mathit{SW} \mathcal{L}_\mathrm{SW}(\phi(\hat{\mathbf{y}}), \phi(\mathbf{t})),
    \label{eqn:loss_selfDZSR}
  \end{equation}
  where $\phi$ denotes the pre-trained VGG-19~\cite{VGGLOSS} network, and we set $\lambda_\mathit{SW}=0.08$.

%
%
\section{Experiments} \label{sec:experiments}
\subsection{Experimental Setup}
  {\textbf{Datasets.}}
  Experiments are conducted on Nikon camera images from DRealSR dataset~\cite{CDC} and the CameraFusion dataset~\cite{DCSR}.
  The training patches of DRealSR have been manually and carefully selected for mitigating the alignment issue, which is laborious and time-consuming.
  Instead, we take the original captured data without manual processing for training, making the whole process fully automated.
  In particular, each scene of the orginal data contains four different focal-length images. 
  We adopt the longest focal-length image as the telephoto input and the shortest focal-length image as the short-focus input, which forms a $\times4$ DZSR dataset.
  There are 163 image pairs for training and 20 images for evaluation.
  In the CameraFusion~\cite{DCSR} dataset, the telephoto and short-focus images are from two lenses with different focal lengths of a smartphone.
  The focal length ratio between the telephoto and short-focus images is $\sim$2.
  Thus, it can constitute a $\times2$ DZSR dataset.
  For this dataset, we use 112 image pairs for training and 12 images for evaluation.

  \noindent\textbf{Data Pre-processing.}
  We first crop the center area of the short-focus image as LR.
  Then the brightness and color matching are employed between the LR and the telephoto images.
  Next, we use PWC-Net~\cite{PWC-Net} to calculate the optical flow between the LR and the telephoto images, and warp the telephoto image.
  The warped telephoto image is used as GT and the center patch of it can be seen as Ref.
  Note that misalignment can still occur after registration, and our SelfDZSR is suggested to learn deep DZSR model while alleviating the adverse effect of misalignment.

  \noindent\textbf{Training Configurations.}
  We augment the training data with random horizontal flip, vertical flip and $90^\circ$ rotation.
  The batch size is 16, and the patch size for LR is 48$\times$48.
  %
  %
  The model is trained with the Adam optimizer~\cite{Adam} by setting $\beta_1=0.9$ and $\beta_2=0.999$ for 400 epochs. 
  The learning rate is initially set to $1\times10^{-4}$ and is decayed to $5\times10^{-5}$ after 200 epochs.
  The experiments are conducted with PyTorch~\cite{PyTorch} on an Nvidia GeForce RTX 2080Ti GPU.

 \noindent\textbf{Evaluation Configurations.}
  There is no GT when inputting the original short-focus and telephoto images directly, so we also use the processed LR and Ref as input and warped telephoto as GT to compare the various methods by rule and line.
  %
  %
  Three common metrics (\ie, PSNR, SSIM~\cite{SSIM} and LPIPS~\cite{LPIPS}) on RGB channels are computed.
  Noted that the scene of the Ref is the same as the center area of LR.
  In addition to calculating the metrics on the full image (marked as \emph{Full-Image}), we also calculate the metrics of the area excluding the center (marked as \emph{Corner-Image}).
  And all patches for visual comparison are selected from the area excluding the center of the whole image. 
  
 \begin{figure}[t]
    \begin{minipage}[t]{\linewidth}
        \centering
        \small
        \subfloat[][\small LR]
        {
            \includegraphics[width=.18\linewidth]{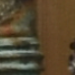}
        }
        \subfloat[][\small MASA~\cite{MASA-SR}]
        {
            \includegraphics[width=.18\linewidth]{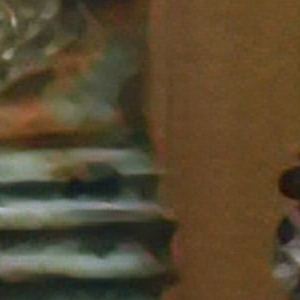}
        }
        \subfloat[][\small DCSR~\cite{DCSR}]
        {
            \includegraphics[width=.18\linewidth]{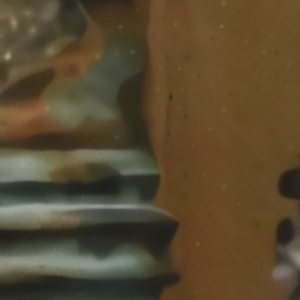}
        }
        \subfloat[][\small SelfDZSR]
        {
            \includegraphics[width=.18\linewidth]{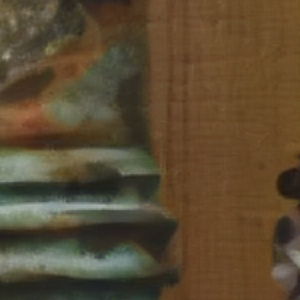}
        }
        \subfloat[][\small GT]
        {
            \includegraphics[width=.18\linewidth]{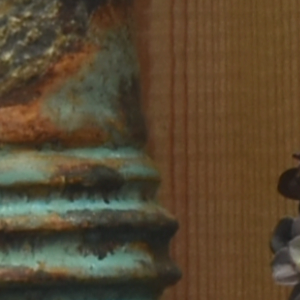}
        }
    \end{minipage}
    \setlength{\abovecaptionskip}{0.5mm}
    \setlength{\belowcaptionskip}{-6.5mm}
    \caption{Visual comparison on Nikon camera images.} 
    \label{fig:Nikon-result} 
\end{figure}
\begin{table}[t!] 
  \scriptsize
  \setlength{\abovecaptionskip}{-2.5mm}
  \caption{Quantitative results on Nikon camera images. Best results are highlighted by {\color{red}red}. The models trained only with $\ell_1$ (or $\ell_2$) loss are marked in gray. RefSR$^\dagger$ represents that the RefSR methods are trained in our self-supervised learning manner.}
  \label{tab:Nikon}
  \centering\noindent
  \centering%
  \begin{center}
    \begin{tabular}{clccc}
      \toprule
      & Method
      & \tabincell{c}{\# Param\\(M)}
      & \tabincell{c}{\emph{Full-Image} \\ {{PSNR}{$\uparrow$} {/} {SSIM}{$\uparrow$} {/} {LPIPS}{$\downarrow$}}}
      & \tabincell{c}{\emph{Corner-Image} \\ {{PSNR}{$\uparrow$} {/} {SSIM}{$\uparrow$} {/} {LPIPS}{$\downarrow$}}} \\
        \midrule
        \multirow{5}{*}{SISR}
        &  \cellcolor{Gray}EDSR~\cite{EDSR}  &  \cellcolor{Gray}43.1 &  \cellcolor{Gray}27.26 / 0.8364 / 0.362 &  \cellcolor{Gray}27.29 / 0.8345 / 0.363 \\ 
        &  \cellcolor{Gray}RCAN~\cite{RCAN}  &  \cellcolor{Gray}15.6 &  \cellcolor{Gray}27.30 / 0.8344 / 0.383 &  \cellcolor{Gray}27.33 / 0.8323 / 0.383 \\ 
        &  \cellcolor{Gray}CDC~\cite{CDC}    &  \cellcolor{Gray}39.9 &  \cellcolor{Gray}27.20 / 0.8306 / 0.412 &  \cellcolor{Gray}27.24 / 0.8283 / 0.412 \\ 
        & BSRGAN~\cite{BSRGAN}            & 16.7 & 26.91 / 0.8151 / 0.279 & 26.96 / 0.8135 / 0.278 \\ 
        & Real-ESRGAN~\cite{Real-ESRGAN}  & 16.7 & 25.96 / 0.8076 / 0.272 & 26.00 / 0.8063 / 0.271 \\ 
        \midrule
        \multirow{10}{*}{RefSR$^\dagger$}  
        &  \cellcolor{Gray}SRNTT-$\ell_2$~\cite{SRNTT}   & \cellcolor{Gray}5.5    &  \cellcolor{Gray}27.30 / 0.8387 / 0.359 &  \cellcolor{Gray}27.33 / 0.8366 / 0.359 \\ 
        & SRNTT~\cite{SRNTT}                 & 5.5    & 27.31 / 0.8242 / 0.286 & 27.35 / 0.8223 / 0.283  \\ 
        &  \cellcolor{Gray}TTSR-$\ell_1$~\cite{TTSR}     &  \cellcolor{Gray}7.3 &  \cellcolor{Gray}25.83 / 0.8272 / 0.369 &  \cellcolor{Gray}25.80 / 0.8259 / 0.369 \\ 
        & TTSR~\cite{TTSR}                   & 7.3 & 25.31 / 0.7719 / 0.282 & 25.27 / 0.7708 / 0.282 \\ 
        &  \cellcolor{Gray}$C^2$-Matching-$\ell_1$~\cite{C2-Matching}  &  \cellcolor{Gray}8.9 &  \cellcolor{Gray}27.19 / 0.8402 / 0.362 &  \cellcolor{Gray}27.23 / 0.8381 / 0.362 \\ 
        & $C^2$-Matching~\cite{C2-Matching}  & 8.9 & 26.79 / 0.8141 / 0.327 & 26.81 / 0.8123 / 0.325 \\
        &  \cellcolor{Gray}MASA-$\ell_1$~\cite{MASA-SR}  &  \cellcolor{Gray}4.0 &  \cellcolor{Gray}27.27 / 0.8372 / 0.339 &  \cellcolor{Gray}27.30 / 0.8352 / 0.339 \\ 
        & MASA~\cite{MASA-SR}                & 4.0 & 27.32 / 0.7640 / 0.273  & 27.37 / 0.7615 / 0.274 \\
        &  \cellcolor{Gray}DCSR-$\ell_1$~\cite{DCSR}    &  \cellcolor{Gray}3.2 &  \cellcolor{Gray}27.73 / 0.8274 / 0.355 &  \cellcolor{Gray}27.72 / 0.8275 / 0.349 \\
        & DCSR~\cite{DCSR}                   & 3.2  & 27.69 / 0.8232 / 0.276 & 27.68 / 0.8232 / 0.272 \\
        \midrule
        \multirow{2}{*}{Ours}  
        &  \cellcolor{Gray}SelfDZSR-$\ell_1$  &  \cellcolor{Gray}3.2  &  \cellcolor{Gray}{\color{red}28.93} / {\color{red}0.8572} / 0.308 &  \cellcolor{Gray}{\color{red}28.67} / {\color{red}0.8457} / 0.328 \\ 
        & SelfDZSR   & 3.2  & 28.67 / 0.8356 / {\color{red}0.219} & 28.42 / 0.8238 / {\color{red}0.231} \\ 
      \bottomrule
    \end{tabular}
    \end{center}
\end{table}  
\setlength{\textfloatsep}{2mm}
\subsection{Results on Nikon Camera}
  %
  We compare results with SISR (\ie, EDSR~\cite{EDSR}, RCAN~\cite{RCAN}, CDC~\cite{CDC}, BSRGAN~\cite{BSRGAN} and Real-ESRGAN~\cite{Real-ESRGAN}) and RefSR (\ie, SRNTT~\cite{SRNTT}, TTSR~\cite{TTSR}, MASA~\cite{MASA-SR}, $C^2$-Matching~\cite{C2-Matching} and DCSR~\cite{DCSR}) methods.
  The results of BSRGAN and Real-ESRGAN are generated via the officially released model, other methods are retrained using our processed data for a fair comparison.
  Among them, RefSR methods are trained in our self-supervised learning manner and each method has two models, obtained by minimizing $\ell_1$ (or $\ell_2$) loss and all loss terms that are used in their papers.

  Benefiting from the implicit alignment of the data pairs and better utilization of Ref information, SelfDZSR exceeds all competing methods on all metrics from Table~\ref{tab:Nikon}.
  As shown in Fig.~\ref{fig:Nikon-result}, our visual result restores much more details.
  More visual results and the evaluation of generalization capacity on other cameras will be given in the suppl.
\begin{table}[t] 
  \scriptsize
  \centering
  \setlength{\abovecaptionskip}{-2mm}
  \caption{Quantitative results on CameraFusion dataset. Best results are highlighted by {\color{red}red}. The models trained only with $\ell_1$ (or $\ell_2$) loss are marked in gray. RefSR$^\dagger$ represents that the RefSR methods are trained in our self-supervised learning manner.}
  \label{tab:Cam}
  \centering\noindent
  \centering%
  \begin{center}
    \begin{tabular}{clcccc}
      \toprule
      & Method
      & \tabincell{c}{\# Params\\(M)}
      & \tabincell{c}{\emph{Full-Image} \\ {{PSNR}{$\uparrow$} {/} {SSIM}{$\uparrow$} {/} {LPIPS}{$\downarrow$}}}
      & \tabincell{c}{\emph{Corner-Image} \\ {{PSNR}{$\uparrow$} {/} {SSIM}{$\uparrow$} {/} {LPIPS}{$\downarrow$}}} \\
        \midrule
        \multirow{5}{*}{SISR}  
        & \cellcolor{Gray}EDSR~\cite{EDSR}   & \cellcolor{Gray}43.1   &  \cellcolor{Gray}25.43 / 0.8041 / 0.356 &  \cellcolor{Gray}25.25 / 0.8007 / 0.349 \\ 
        & \cellcolor{Gray}RCAN~\cite{RCAN} & \cellcolor{Gray}15.6 &  \cellcolor{Gray}25.31 / 0.8034 / 0.355 &  \cellcolor{Gray}25.14 / 0.8004 / 0.349 \\ 
        & \cellcolor{Gray}CDC~\cite{CDC}  & \cellcolor{Gray}39.9 &  \cellcolor{Gray}24.31 / 0.7811 / 0.380 &  \cellcolor{Gray}24.11 / 0.7771 / 0.374 \\ 
        & BSRGAN~\cite{BSRGAN}  & 16.7 &  25.09 / 0.7779 / 0.272 &  24.92 / 0.7749 / 0.266 \\ 
        & Real-ESRGAN~\cite{Real-ESRGAN}   & 16.7 &  25.13 / 0.7788 / 0.261 &  24.90 / 0.7755 / 0.255 \\ 
        \midrule
        \multirow{10}{*}{RefSR$^\dagger$}  
        & \cellcolor{Gray}SRNTT-$\ell_2$~\cite{SRNTT}   &  \cellcolor{Gray}5.5    & \cellcolor{Gray}24.78 / 0.7781 / 0.333   & \cellcolor{Gray}24.49 / 0.7737 / 0.331 \\ 
        & SRNTT~\cite{SRNTT}  & 5.5    & 23.69 / 0.7740 / 0.230   & 23.38 / 0.7700 / 0.229 \\ 
        & \cellcolor{Gray}TTSR-$\ell_1$~\cite{TTSR}      &  \cellcolor{Gray}7.3    & \cellcolor{Gray}24.42 / 0.7937 / 0.375  & \cellcolor{Gray}24.12 / 0.7901 / 0.372 \\ 
        & TTSR~\cite{TTSR}   &  7.3    & 23.05 / 0.7879 / 0.303  & 22.74 / 0.7854 / 0.300 \\   
        & \cellcolor{Gray}$C^2$-Matching-$\ell_1$~\cite{C2-Matching}            & \cellcolor{Gray}8.9  & \cellcolor{Gray}25.24 / 0.7992 / 0.346 & \cellcolor{Gray}25.07 / 0.7971 / 0.340 \\ 
        & $C^2$-Matching~\cite{C2-Matching}  & 8.9  & 24.18 / 0.7252 / 0.252 & 24.06 / 0.7254 / 0.245 \\
        & \cellcolor{Gray}MASA-$\ell_1$~\cite{MASA-SR}   & \cellcolor{Gray}4.0  & \cellcolor{Gray}25.78 / 0.8063 / 0.335 & \cellcolor{Gray}25.52 / 0.8026 / 0.331 \\ 
        & MASA~\cite{MASA-SR}  & 4.0   &25.42 / 0.7543 / 0.194 & 25.27 / 0.7524 / 0.190 \\ 
        & \cellcolor{Gray}DCSR-$\ell_1$~\cite{DCSR}    & \cellcolor{Gray}3.2  & \cellcolor{Gray}25.80 / 0.7974 / 0.300 & \cellcolor{Gray}25.48 / 0.7932 / 0.298 \\
        & DCSR~\cite{DCSR}                   & 3.2  & 25.51 / 0.7890 / 0.209 & 25.20 / 0.7847 / 0.211 \\
        \midrule
        \multirow{2}{*}{Ours}  
        & \cellcolor{Gray}SelfDZSR-$\ell_1$  & \cellcolor{Gray}3.2  & \cellcolor{Gray}{\color{red}26.35} / {\color{red}0.8276} / 0.262 & \cellcolor{Gray}{\color{red}25.67} / {\color{red}0.8040} / 0.292 \\ 
        & SelfDZSR   & 3.2  & 26.03 / 0.8008 / {\color{red}0.158} & 25.37 / 0.7740 / {\color{red}0.174} \\ 
      \bottomrule
    \end{tabular}
    \end{center}
\end{table}
\subsection{Results on CameraFusion Dataset}
  %
  Different from the Nikon camera images which scale factor of SR is $\times$4, it is $\times$2 for the CameraFusion dataset.
  Other settings of experiments on the CameraFusion dataset are the same as those on the Nikon camera images.
  Table~\ref{tab:Cam} shows the comparison of quantitative results on the CameraFusion dataset, and we still achieve the best results among SISR and RefSR methods.
  The qualitative comparison will be given in the suppl.
\section{Ablation Study}
\label{sec:ablation}
  In this section, we conduct ablation experiments for assessing the effect of self-supervised learning, auxiliary-LR for alignment and AdaSTN.
  Unless otherwise stated, experiments are carried out on the Nikon camera images~\cite{CDC}, and the metrics are evaluated on full images.
\subsection{Effect of Self-Supervised Learning}
  In order to verify the effectiveness of our proposed self-supervised approach, we conduct experiments on different training strategies.
  First, we remove the auxiliary-LR generator and AdaSTN in SelfDZSR.
  Then we replace the real LR and auxiliary-LR images with the bicubic downsampling HR image, and retrain the network.
  Finally, for a fair comparison, we take the self-supervised real-image adaptation (SRA)~\cite{DCSR} and our self-supervised strategy to fine-tune the above model, respectively.
  As can be seen from Table~\ref{tab:ablation_self}, when evaluating on real-world images, our proposed self-supervised method achieves the better results. 
  The PSNR metric is 0.27 dB higher than the model based on SRA fine-tuning.
  And our visual result is sharper and clearer. 
  
  Moreover, for the CameraFusion dataset, DCSR~\cite{DCSR} model trained by our self-supervised approach obtains a 0.31 dB PSNR gain in comparison to the officially released model.
  In a word, it can be seen that even if the misalignment between LR and GT is not handled, our self-supervised method is still better than SRA~\cite{DCSR} strategy.
\begin{table}[t!] 
  \small
  \setlength{\abovecaptionskip}{-2mm}
  \setlength{\belowcaptionskip}{-2mm}
  \caption{Ablation study on training strategies.}
  \label{tab:ablation_self}
  \centering\noindent
  \centering%
  \begin{center}
    \begin{tabular}{cccc}
      \toprule  
      {Training Strategy\quad} & {Bicubic Degradation\quad} & {SRA Fine-tuning~\cite{DCSR}\quad} & {Our Fine-tuning} \\
       \midrule
       {{PSNR}{$\uparrow$} / {LPIPS}{$\downarrow$}\quad}  & {28.07 / 0.398\quad}  & {28.26 / 0.278\quad}  & {28.53 / 0.223} \\ 
      \bottomrule
    \end{tabular}
    \end{center}
\end{table}
\begin{table}[t] 
  \small
  \setlength{\abovecaptionskip}{-2mm}
  \setlength{\belowcaptionskip}{-4mm}
  \caption{Ablation study on alignment methods. Data pairs are pre-aligned by~\cite{PWC-Net}. '\texttimes' represents replacing auxiliary-LR with LR.}
  \label{tab:ablation_alignment}
  \centering\noindent
  \centering%
  \begin{center}
    \begin{tabular}{ccc}
      \toprule  
      \tabincell{c}{Align LR with auxiliary-LR \qquad} & 
      \tabincell{c}{Align Ref with auxiliary-LR \qquad}  & 
      {{PSNR}{$\uparrow$ / LPIPS}{$\downarrow$}} \\
        \midrule
       \texttimes  \qquad & \texttimes   \qquad & 28.48 / 0.222 \\ 
       \checkmark  \qquad & \texttimes   \qquad & 28.67 / 0.224 \\ 
       \texttimes  \qquad & \checkmark   \qquad & 28.61 / 0.220 \\ 
       \checkmark  \qquad & \checkmark   \qquad & 28.67 / 0.219 \\ 
      \bottomrule
    \end{tabular}
    \end{center}
\end{table}
\begin{table}[t] 
  \small
  \setlength{\abovecaptionskip}{-2mm}
  \setlength{\belowcaptionskip}{-2mm}
  \caption{Ablation study on loss terms of auxiliary-LR generator. $\lambda_\mathit{p}$ denotes the coefficient of position preserving loss in Eqn.~(\ref{eqn:loss_d}).}
  \label{tab:ablation_loss}
  \centering\noindent
  \centering%
  \begin{center}
    \begin{tabular}{ccccc}
      \toprule  
      {$\lambda_\mathit{p}$ \quad} &  0 \qquad & 1 \qquad & 100 \qquad & 10000 \\
      \midrule
      {{PSNR}{$\uparrow$ / LPIPS}{$\downarrow$} \quad} & 28.22 / 0.225 \qquad & 28.47 / 0.221 \qquad & 28.67 / 0.219 \qquad & 28.28 / 0.222 \\
      \bottomrule
    \end{tabular}
    \end{center}
\end{table}
\begin{table}[t!] 
  \small
   \setlength{\abovecaptionskip}{-2mm}
  \setlength{\belowcaptionskip}{-4mm}
  \caption{Ablation study on AdaSTN.}
  \label{tab:ablation_AdaSTN}
  \centering\noindent
  \centering%
  \begin{center}
    \begin{tabular}{lc}
      \toprule
      Method & {{PSNR}{$\uparrow$} / {LPIPS}{$\downarrow$}} \\
      \midrule
      Baseline & 28.48 / 0.222 \\ 
      Baseline + Deformable Conv~\cite{DefConv}  & 28.52 / 0.225 \\ 
      Baseline + STN~\cite{STN} & 28.57 / 0.219 \\ 
      Baseline + AdaSTN & 28.67 / 0.219 \\ 
      \bottomrule
    \end{tabular}
    \end{center}
\end{table}

%
%
\subsection{Effect of Auxiliary-LR for Alignment}
  In order to evaluate the effect of our alignment methods, we experiment on the role of auxiliary-LR in alignment by replacing auxiliary-LR with LR during the training, which corresponds to Sec.~\ref{sec:alignment_gt} and Sec.~\ref{sec:alignment_ref}.
  We consider the model that does not leverage auxiliary-LR as the baseline, which also means that the training data is only pre-aligned by optical flow~\cite{PWC-Net}.
  When aligning LR with auxiliary-LR only, the PSNR increases by 0.19 dB against the baseline, as shown in Table~\ref{tab:ablation_alignment}.
  Coupled with the alignment between Ref and auxiliary-LR, better quantitative results can be further attained.

  In addition, we conduct an experiment that replaces auxiliary-LR with bicubically down-sampled GT, and PSNR drops by 0.47 dB, LPIPS gets worse by 0.165.
  The result shows the auxiliary-LR generator is necessary and effective.
  We also conduct experiments on different coefficients (\ie, $\lambda_\mathit{p}$) of position preserving loss, as shown in Table~\ref{tab:ablation_loss}.
  %
  %
  In order to bring auxiliary-LR into play better on alignment and obtain better SR performance, we take a trade-off between content preserving loss and position preserving loss, and set $\lambda_\mathit{p}$ to 100.
  %

\subsection{Effect of AdaSTN}
  We regard the model only using flow-based alignment~\cite{PWC-Net} as the baseline.
  We modify the proposed AdaSTN to deformable convolution~\cite{DefConv} and STN~\cite{STN} to verify the effect of AdaSTN.
  For deformable convolution, instead of calculating the offset by estimating the affine transformation matrix and vector according to Eqn.~(\ref{eqn:offset}), we estimate the offset directly.
  For STN, we replace the pixel-level offset with a global affine transformation.
  The PSNR gain of taking AdaSTN is 0.15 dB compared with deformable convolution and 0.1 dB compared with STN, as shown in Table~\ref{tab:ablation_AdaSTN}.
  %
  %
  In addition, for the image with the size of 1445$\times$945, setting $\mathbf{P} = \mathbf{0}$ (see Eqn.~(\ref{eqn:offset})) for AdaSTNs increases the average inference speed by $\sim$0.2 seconds without performance dropping.

\section{Conclusion}
  Real-world image super-resolution from dual zoomed observations (DZSR) is an emerging topic, which aims to super-resolve the short focal length image with the reference of telephoto image.
  To circumvent the problem that ground-truth is unavailable, we present an effective self-supervised learning method, named SelfDZSR.
  To mitigate the adverse effect of image misalignment during training, the auxiliary-LR that is aligned with GT is generated to guide the alignment of LR and Ref towards GT.
  And with the help of auxiliary-LR, we propose adaptive spatial transformer networks (AdaSTN) to align LR with GT.
  Experiments show that our proposed method can achieve better performance against the state-of-the-art methods both quantitatively and qualitatively.
  %
\section*{Acknowledgement}
  This work was supported by Alibaba Group through Alibaba Innovative Research Program, the Major Key Project of Peng Cheng Laboratory (PCL2021A12), and the National Natural Science Foundation of China (NSFC) under Grants No.s 61872118 and U19A2073.

\clearpage
\appendix

\renewcommand{\thesection}{\Alph{section}}
\renewcommand{\thetable}{\Alph{table}}
\renewcommand{\thefigure}{\Alph{figure}}
\renewcommand{\thealgorithm}{\Alph{algorithm}}

\setcounter{section}{0}
\setcounter{table}{0}
\setcounter{figure}{0}
\setcounter{algorithm}{0}

\section*{\centering{\Large Supplementary Material\\[30pt]}}

\section{Content}
  The content of this supplementary material involves:
  \begin{itemize}
    \item  Synthetic noise for auxiliary-LR in Sec.~\ref{sec:noise}.  %
    \item  Visual results of auxiliary-LR in Sec.~\ref{sec:vis-lr}.  %
    \item Network structure of restoration module in Sec.~\ref{sec:restoration-supp}. %
    \item Sliced Wasserstein (SW) loss in Sec.~\ref{sec:swloss}.  %
    \item Comparison of  \#FLOPs in Sec.~\ref{sec:flops}.  %
    \item Evaluation of generalization performance on other cameras in Sec.~\ref{sec:generalizability}.  %
    \item Additional visual comparison on Nikon camera and CameraFusion dataset in Sec.~\ref{sec:visual}. %
  \end{itemize}

\section{Synthetic Noise for Auxiliary-LR} \label{sec:noise}
  Noise in real-world images is common, but complex and various.
  In order to bridge the gap between auxiliary-LR and LR as much as possible, we need to add noise to the output of the auxiliary-LR generator network that only simulates the blurring and down-sampling processes.
  Gaussian noise is a natural choice, but much different from real-world image noise.
  Inspired by BSRGAN~\cite{BSRGAN}, we also add JPEG compression noise.
  %
  The variance of Gaussian noise is uniformly sampled from 5/255 to 30/255, and the JPEG quality factor is uniformly chosen from 60 to 95. 
  Simultaneously, the order of adding three kinds of noise is stochastic.
 
\section{Visual Results of Auxiliary-LR} \label{sec:vis-lr}
\begin{figure*}[t]
    \centering
    \begin{overpic}
      [width=\linewidth]{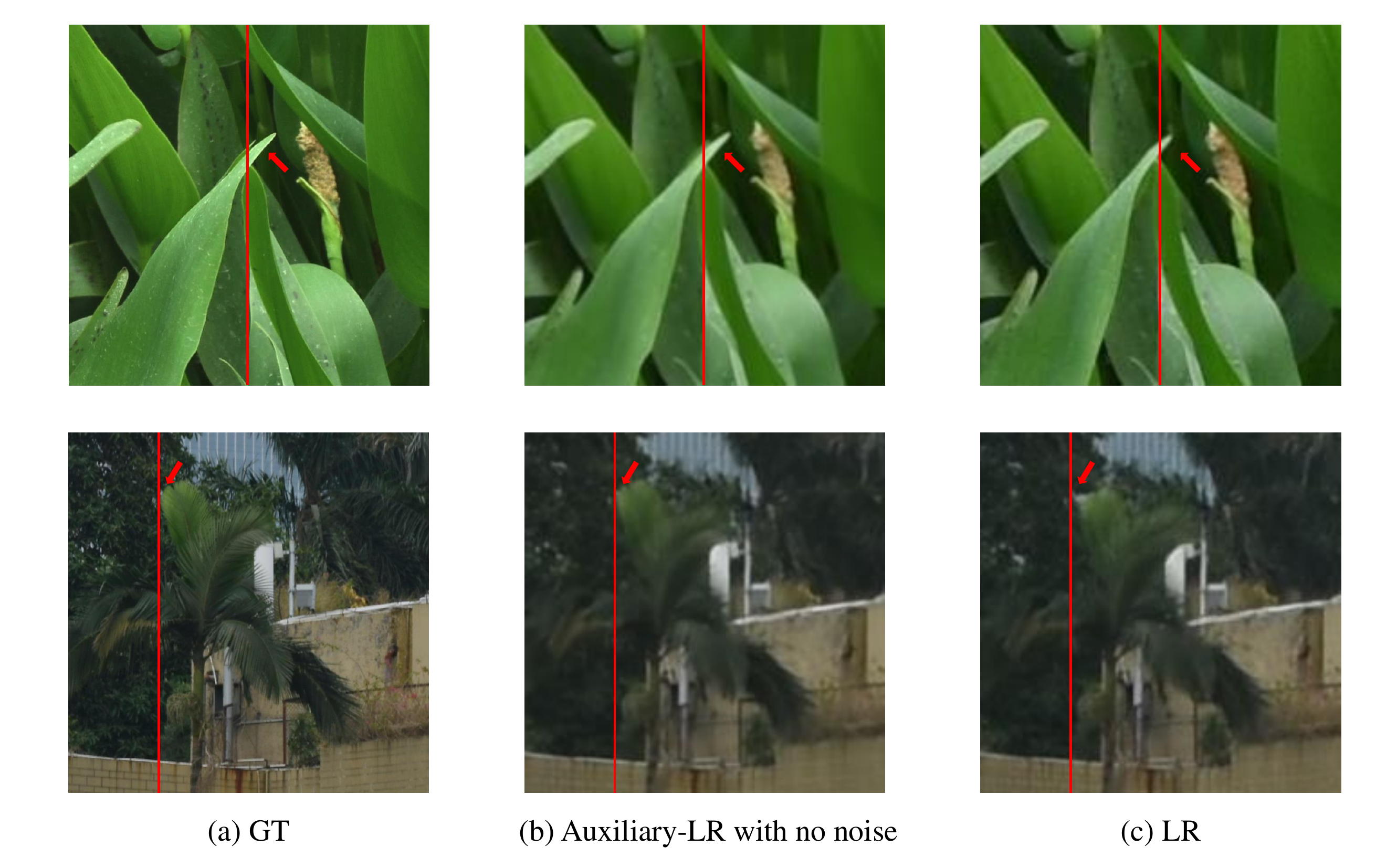}
    \end{overpic}
    \vspace{-2mm}
    \caption{Visual Results of auxiliary-LR with no noise. The auxiliary-LR has similar contents as LR and it is aligned with GT.}
    \label{fig:misalignment}
\end{figure*}

Since the noise type and intensity of auxiliary-LR are random, for the convenience of display and comparison, we show the auxiliary-LR images without adding synthetic noise in Fig.~\ref{fig:misalignment}(b).
And the corresponding GT and LR images are shown in Fig.~\ref{fig:misalignment}(a) and Fig.~\ref{fig:misalignment}(c), respectively.
The red lines and arrows in the same row are in the same position relative to the image.
It can be seen that the auxiliary-LR has similar contents as LR and it is aligned with GT.
It indicates that the function of the auxiliary-LR generator is guaranteed.

\begin{figure*}[t!]
    \centering
    \vspace{-3mm}
    \begin{overpic}
      [width=\linewidth]{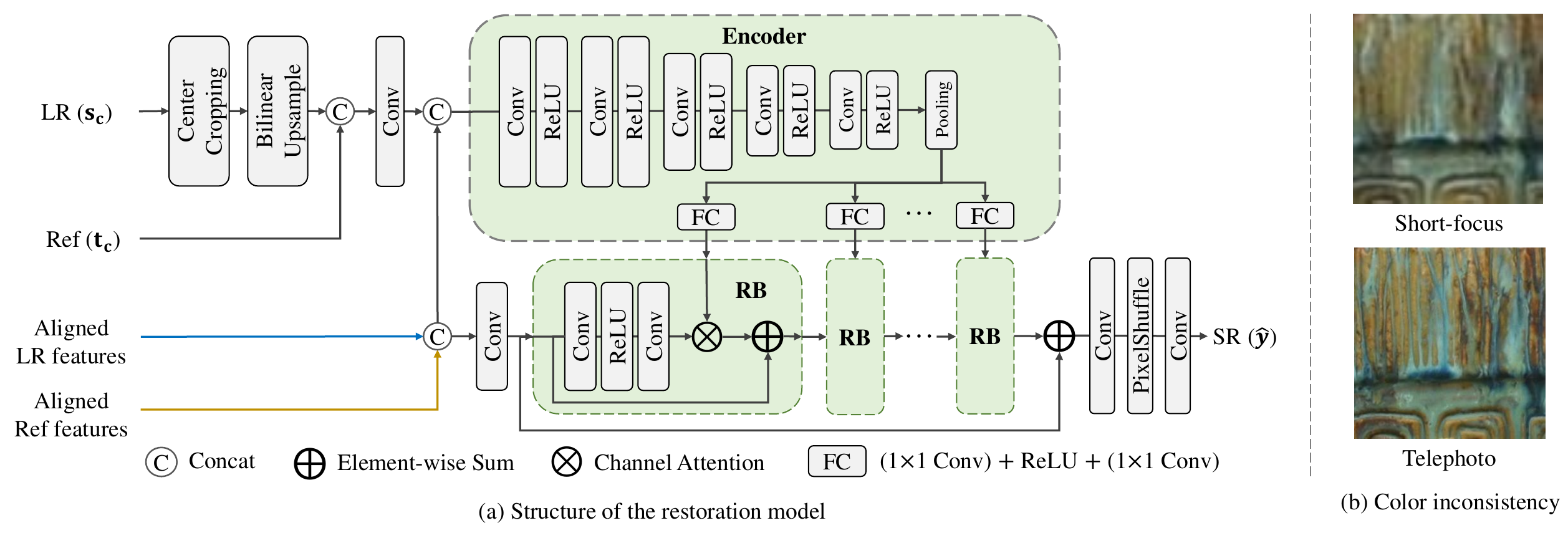}
    \end{overpic}
    \vspace{-2mm}
    \caption{Structure of the restoration model and color inconsistency. (a) Detailed structure of the restoration model. `RB' denotes the residual block.  (b) Color inconsistency between short-focus and telephoto images.
    }
    \label{fig:restore}
    \vspace{6mm}
\end{figure*}

\section{Restoration Module} \label{sec:restoration-supp}
  %
  Fig.~\ref{fig:restore}(a) shows the detailed structure of the restoration module.
  First, the aligned LR and aligned Ref features are concatenated and fed into the backbone, which consists of 16 residual blocks~\cite{EDSR}.
  Then the concatenated features are input into an encoder for generating vectors that modulate the features of each residual block.
  Simultaneously, the original Ref image and the central area of the LR image are utilized to enrich the input features of the encoder for better modulation.
  This modulation can also be regarded as a kind of channel attention on the features of residual block.
  And it is beneficial to relieve the color inconsistency (see Fig.~\ref{fig:restore}(b)) between the short focal length and telephoto images in the real world.

\begin{algorithm*}[t]
  \caption{Pseudocode of SW loss}
  \label{fig:pseudocode}
  \begin{algorithmic}[1]
    \Require
      $\mathbf{U}\in\mathbb{R}^{C \times H \times W}$: VGG features of output image;  $\mathbf{V}\in\mathbb{R}^{C \times H \times W}$: VGG features of target image;
      $\mathbf{M}\in\mathbb{R}^{C' \times C}$: random projection matrix;
    \Ensure
      $\mathcal{L}_\mathrm{SW}(\mathbf{U},\mathbf{V})$: the value of SW loss;
      \State flatten features $\mathbf{U}$ and $\mathbf{V}$ to $\mathbf{U_f}(\in\mathbb{R}^{C \times HW})$ and $\mathbf{V_f}(\in\mathbb{R}^{C \times HW})$, respectively;
      \State project the features onto $C'$ directions: $\mathbf{U_p} = \mathbf{M} \mathbf{U_f}$, $\mathbf{V_p} = \mathbf{M} \mathbf{V_f}$;
      \State sort projections for each direction: $\mathbf{U_s} = \mathbf{Sort}(\mathbf{U_p}$, dim=1), $\mathbf{V_s} = \mathbf{Sort}(\mathbf{V_p}$, dim=1);
      \State $\mathcal{L}_\mathrm{SW}(\mathbf{U},\mathbf{V}) = \|\mathbf{U_s}-\mathbf{V_s}\|_1 $
  \end{algorithmic}
\end{algorithm*}

\begin{table}[t!] 
  \small
  \vspace{-2mm}
  \centering
  \caption{Quantitative results of SelfSZSR using different loss terms.}
  \label{tab:loss}
  \centering\noindent
  \centering%
  \begin{center}
    \begin{tabular}{lcc}
      \toprule
      Loss Terms
      & {PSNR}{$\uparrow$} 
      & {LPIPS}{$\downarrow$} \\
        \hline
        $\ell_1$  & 28.93 & 0.308 \\ 
        $\ell_1$ + SW  & 28.67 & 0.219 \\ 
        $\ell_1$ + Perceptual + Adversarial   & 28.45 & 0.216 \\
      \bottomrule  
    \end{tabular}
    \end{center}
\end{table}

\section{Sliced Wasserstein Loss} \label{sec:swloss}
  The algorithm of SW loss is given in Alg.~\ref{fig:pseudocode}.
  We first obtain the 1-dimensional representation of 2-dimensional VGG~\cite{VGGLOSS} features through random linear projection.
  Then we calculate the Wasserstein distance between the output and the target 1-dimensional probability distributions, which is defined as the element-wise $\ell_1$ distance over sorted 1-dimensional distributions.

  Most reference-based image SR (RefSR) methods~\cite{SRNTT,TTSR,MASA-SR,C2-Matching} adopt the perceptual loss and adversarial loss~\cite{GAN} for more realistic results.
  For a fair comparison, here we also train proposed model (SelfDZSR) using a combination of $\ell_1$ reconstruction loss, perceptual loss and adversarial loss based on Relativistic GAN~\cite{RelateGAN}.
  The quantitative results are shown in Table~\ref{tab:loss}.
  It can be seen that the model trained by SW loss obtains a 0.22 dB PSNR gain than that by adversarial loss, while the gap of LPIPS metric is small.
  Nevertheless, benefiting from the proposed implicit alignment and better utilization of Ref information, the SelfDZSR model by adversarial loss still achieves better performance than other RefSR methods.

\section{Comparison of \#FLOPs} \label{sec:flops}
The cost of calculating similarity between LR and Ref occupies a large part of the computational cost of previous RefSR methods.
We calculate cosine similarity between $\times4$ down-sampled Ref and $\times4$ down-sampled LR features, and find that its performance is close to that of computing similarity at original image size.
By virtue of the faster similarity calculation and more lightweight restoration model, our method has lower FLOPs in comparison to both SISR and RefSR methods, as shown in Table~\ref{tab:flops}.
\begin{table}[t] 
  \centering
  \caption{\#FLOPs comparison of SISR and RefSR methods. The \#FLOPs is measured under the setting of $\times4$  super-resolving LR image to  $1280\times720$ resolution. For RefSR methods, the Ref image has the same size with LR.} %
  \label{tab:flops}
  \centering\noindent
  \centering%
  \begin{center}
    \begin{tabular}{ccc}
      \toprule
      & Method & \#FLOPs (G)  \\
      \midrule
      \multirow{5}{*}{SISR}
      & EDSR~[{\color{green}25}]   & 5792 \\
      & RCAN~[{\color{green}52}]   & 1838 \\
      & CDC~[{\color{green}42}]   & 1626  \\
      & BSRGAN~[{\color{green}48}]   & 2068 \\
      & Real-ESRGAN~[{\color{green}39}]   & 2068  \\
      \midrule
      \multirow{5}{*}{RefSR}  
      & SRNTT~[{\color{green}53}]   & 3568 \\ 
      & TTSR~[{\color{green}47}]   & 2468 \\ 
      & $C^2$-Matching~[{\color{green}19}]   & 1968 \\ 
      & MASA~[{\color{green}27}]   & 1984 \\ 
      & DCSR~[{\color{green}38}]   & 836 \\ 
      \midrule
      Ours & SelfDZSR  & 384 \\ 
      \bottomrule  
    \end{tabular}
    \end{center}
\end{table}

\section{Evaluation of Generalization Performance on Other Cameras} \label{sec:generalizability}
  Here we evaluate the generalization performance of models on other four cameras (\ie, Canon, Olympus, Panasonic and Sony) from DRealSR dataset~\cite{CDC}.
  We compare results with SISR (\ie, EDSR~\cite{EDSR}, RCAN~\cite{RCAN}, CDC~\cite{CDC}, BSRGAN~\cite{BSRGAN} and Real-ESRGAN~\cite{Real-ESRGAN}) and RefSR (\ie, SRNTT~\cite{SRNTT}, TTSR~\cite{TTSR}, MASA~\cite{MASA-SR}, $C^2$-Matching~\cite{C2-Matching} and DCSR~\cite{DCSR}) methods.
  Among them, the results of BSRGAN and Real-ESRGAN are generated via the officially released model while other methods are trained on Nikon camera images, as mentioned in the main text of the submission.  %
  
\begin{figure*}
    \begin{minipage}[ht]{\linewidth}
        \centering
        \subfloat[][Short-focus]
        {
            \includegraphics[height=.142\linewidth, width=.213\linewidth]{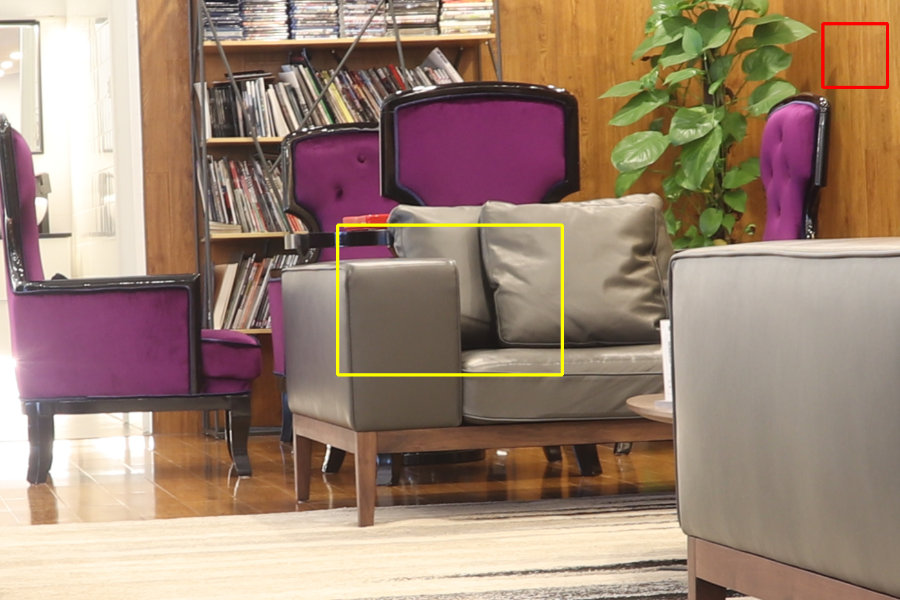}
        }
        \subfloat[][LR]
        {
            \includegraphics[width=.142\linewidth]{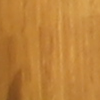}
        }
        \subfloat[][\cite{BSRGAN}]
        {
            \includegraphics[width=.142\linewidth]{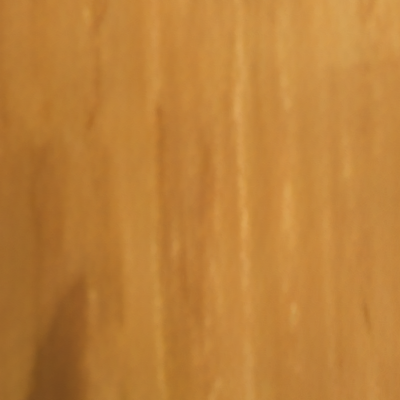}
        }
        \subfloat[][\cite{Real-ESRGAN}]
        {
            \includegraphics[width=.142\linewidth]{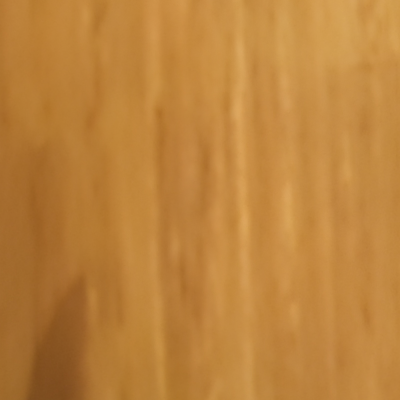}
        }
        \subfloat[][\cite{SRNTT}]
        {
            \includegraphics[width=.142\linewidth]{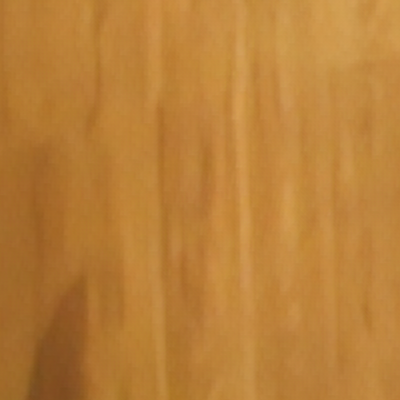}
        }
        \subfloat[][\cite{TTSR}]
        {
            \includegraphics[width=.142\linewidth]{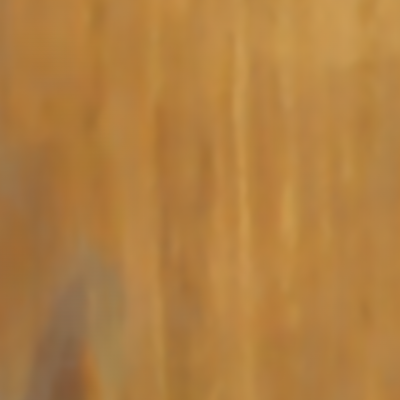}
        }

        \subfloat[][Telephoto]
        {
            \includegraphics[height=.142\linewidth, width=.213\linewidth]{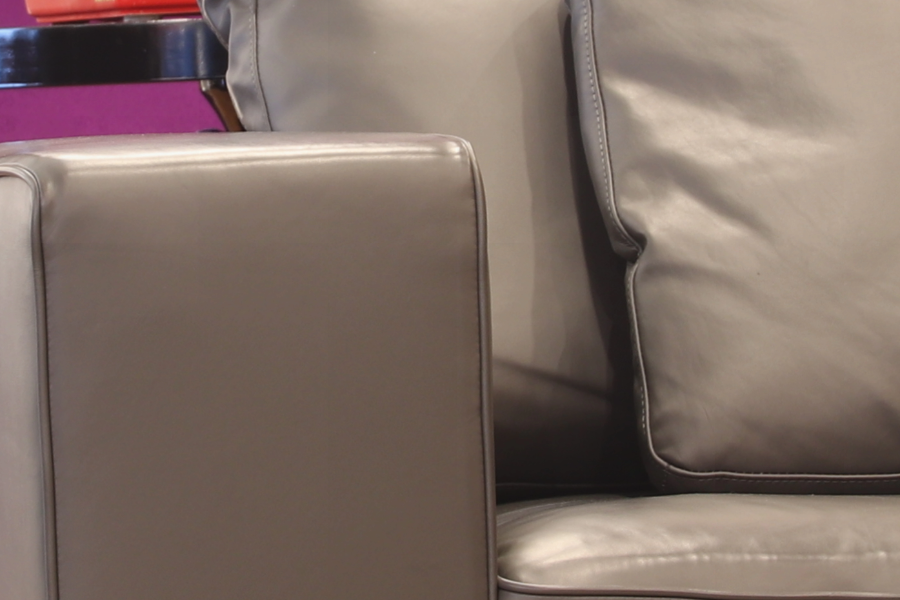}
        }
        \subfloat[][\cite{C2-Matching}]
        {
            \includegraphics[width=.142\linewidth]{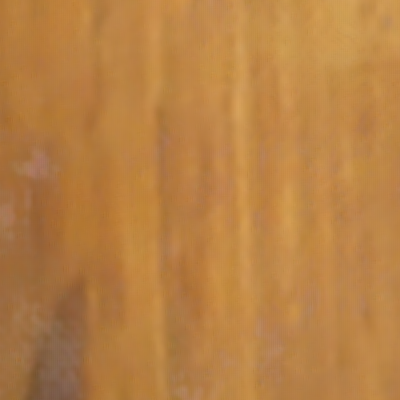}
        }
        \subfloat[][\cite{MASA-SR}]
        {
            \includegraphics[width=.142\linewidth]{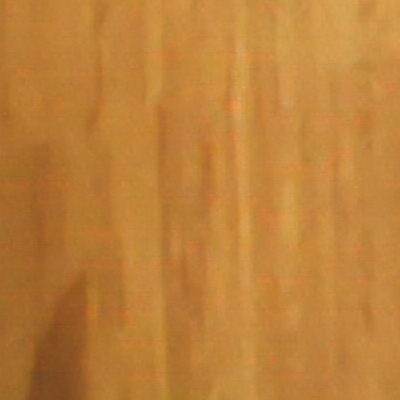}
        }
        \subfloat[][\cite{DCSR}]
        {
            \includegraphics[width=.142\linewidth]{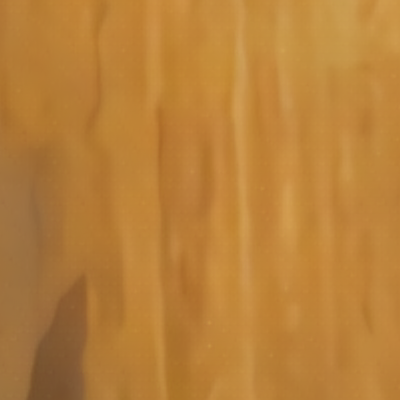}
        }
        \subfloat[][SelfDZSR]
        {
            \includegraphics[width=.142\linewidth]{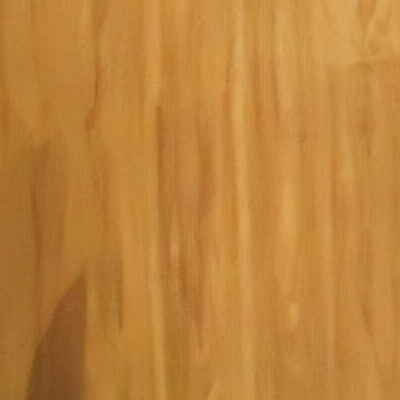}
        }
        \subfloat[][GT]
        {
            \includegraphics[width=.142\linewidth]{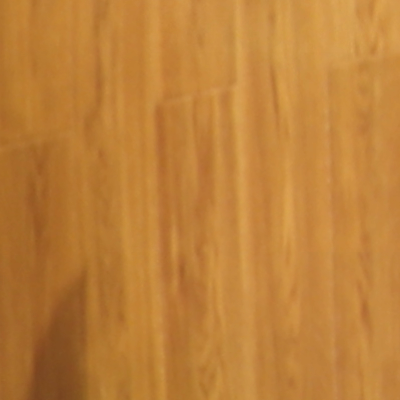}
        }
        

        \subfloat[][Short-focus]
        {
            \includegraphics[height=.142\linewidth, width=.213\linewidth]{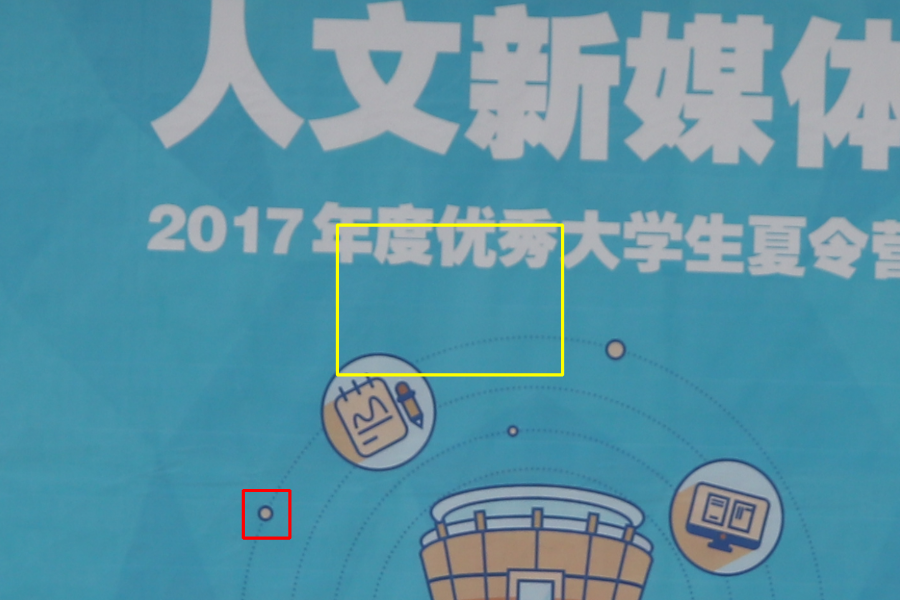}
        }
        \subfloat[][LR]
        {
            \includegraphics[width=.142\linewidth]{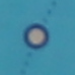}
        }
        \subfloat[][\cite{BSRGAN}]
        {
            \includegraphics[width=.142\linewidth]{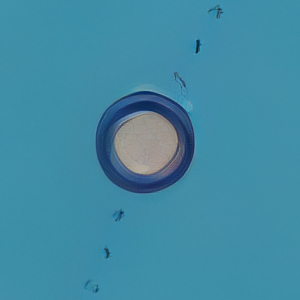}
        }
        \subfloat[][\cite{Real-ESRGAN}]
        {
            \includegraphics[width=.142\linewidth]{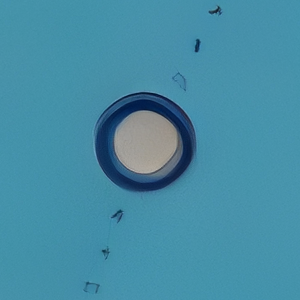}
        }
        \subfloat[][\cite{SRNTT}]
        {
            \includegraphics[width=.142\linewidth]{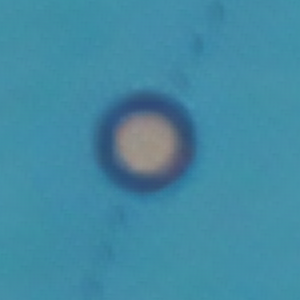}
        }
        \subfloat[][\cite{TTSR}]
        {
            \includegraphics[width=.142\linewidth]{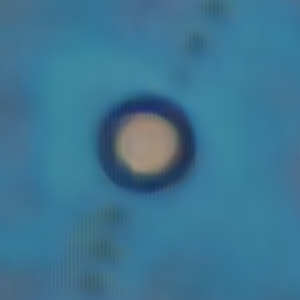}
        }

        \subfloat[][Telephoto]
        {
            \includegraphics[height=.142\linewidth, width=.213\linewidth]{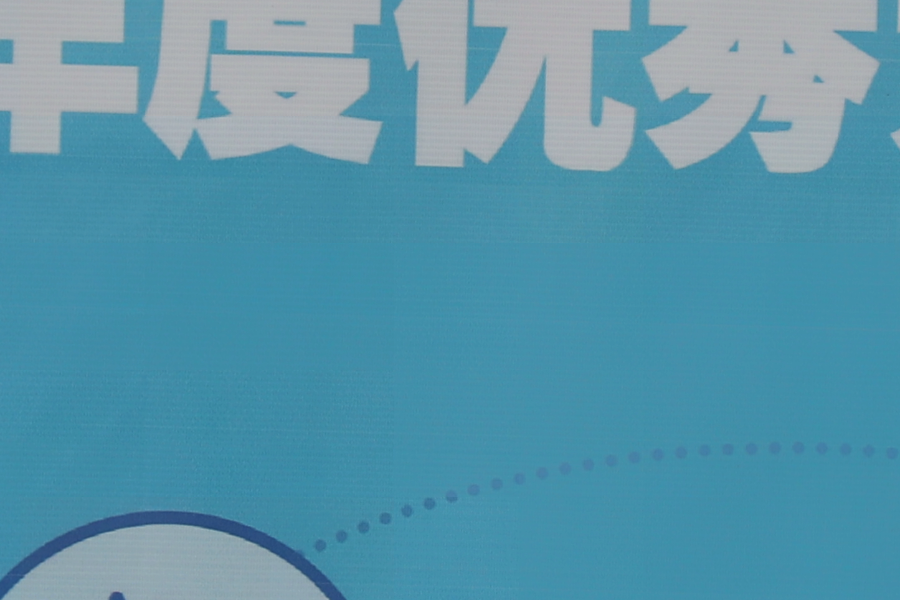}
        }
        \subfloat[][\cite{C2-Matching}]
        {
            \includegraphics[width=.142\linewidth]{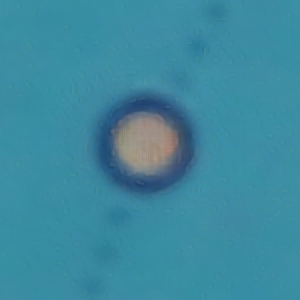}
        }
        \subfloat[][\cite{MASA-SR}]
        {
            \includegraphics[width=.142\linewidth]{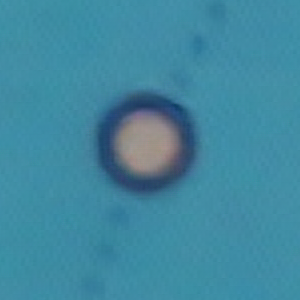}
        }
        \subfloat[][\cite{DCSR}]
        {
            \includegraphics[width=.142\linewidth]{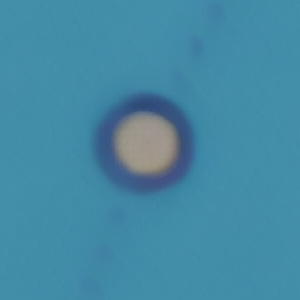}
        }
        \subfloat[][SelfDZSR]
        {
            \includegraphics[width=.142\linewidth]{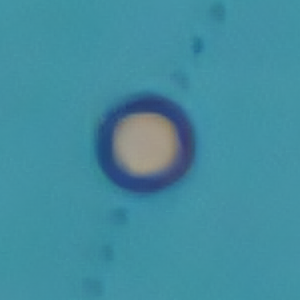}
        }
        \subfloat[][GT]
        {
            \includegraphics[width=.142\linewidth]{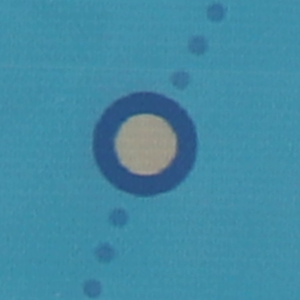}
        }
    \end{minipage}
    \vspace{-1mm}
    \caption{Visual comparison on \textbf{Canon} camera. In the short-focus image, the yellow box indicates the overlapped scene with the telephoto image, while the red box represents the selected LR patch. Our result in sub-figure (k) restores more fine-scale textures, and that in sub-figure (w) is clearer and more photo-realistic.}
    \label{fig:Canon} 
\end{figure*}

\begin{figure*}
    \begin{minipage}[ht]{\linewidth}
        \centering
        \subfloat[][Short-focus]
        {
            \includegraphics[height=.142\linewidth, width=.213\linewidth]{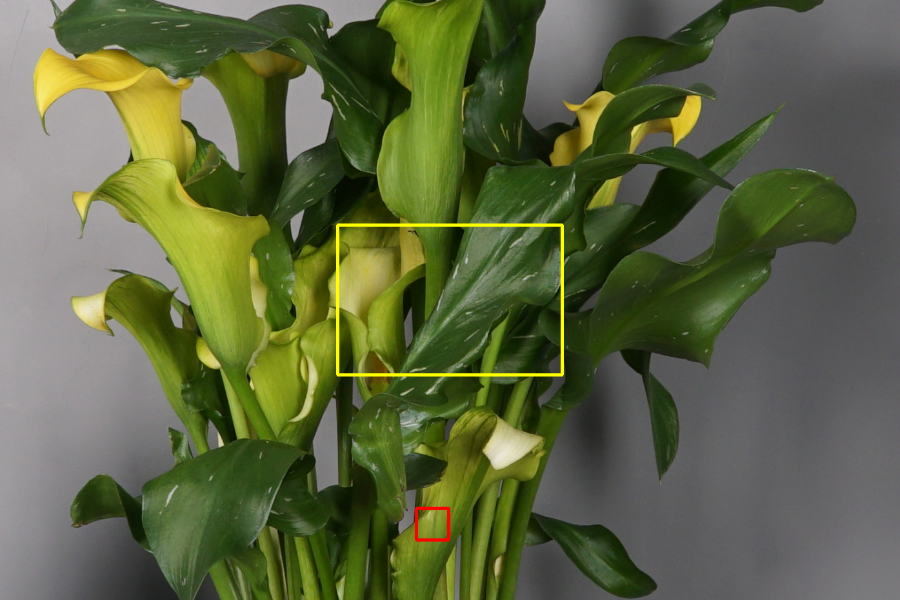}
        }
        \subfloat[][LR]
        {
            \includegraphics[width=.142\linewidth]{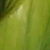}
        }
        \subfloat[][\cite{BSRGAN}]
        {
            \includegraphics[width=.142\linewidth]{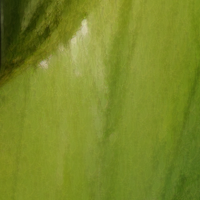}
        }
        \subfloat[][\cite{Real-ESRGAN}]
        {
            \includegraphics[width=.142\linewidth]{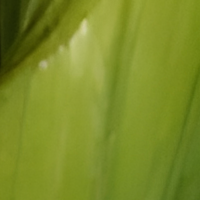}
        }
        \subfloat[][\cite{SRNTT}]
        {
            \includegraphics[width=.142\linewidth]{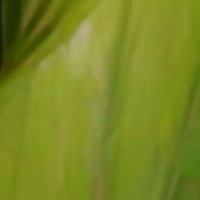}
        }
        \subfloat[][\cite{TTSR}]
        {
            \includegraphics[width=.142\linewidth]{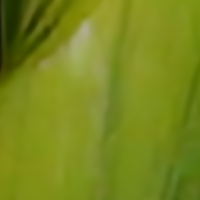}
        }
        
        \subfloat[][Telephoto]
        {
            \includegraphics[height=.142\linewidth, width=.213\linewidth]{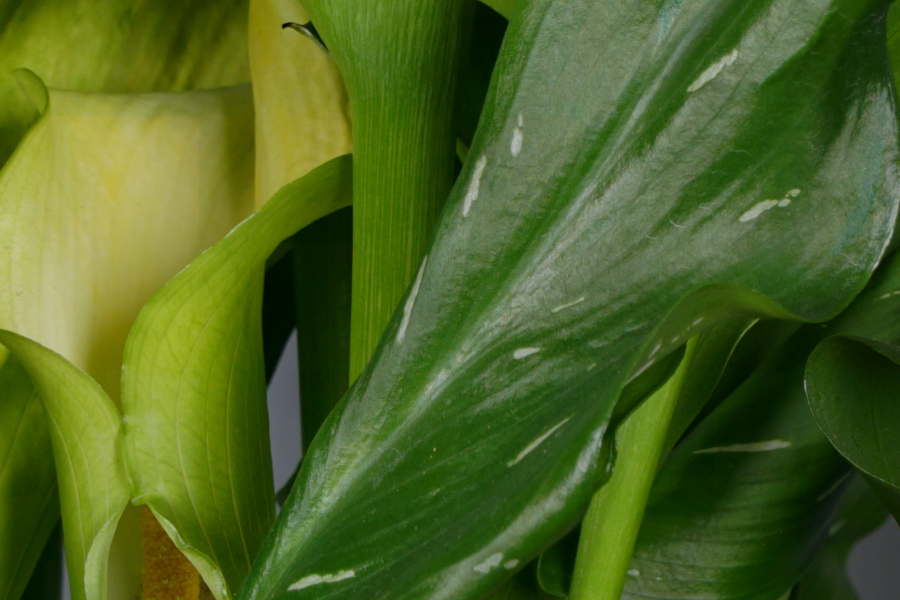}
        }
        \subfloat[][~\cite{C2-Matching}]
        {
            \includegraphics[width=.142\linewidth]{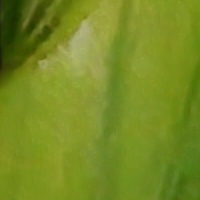}
        }
        \subfloat[][\cite{MASA-SR}]
        {
            \includegraphics[width=.142\linewidth]{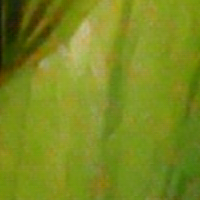}
        }
        \subfloat[][\cite{DCSR}]
        {
            \includegraphics[width=.142\linewidth]{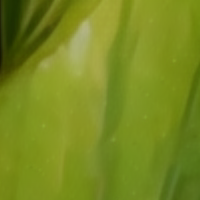}
        }
        \subfloat[][SelfDZSR]
        {
            \includegraphics[width=.142\linewidth]{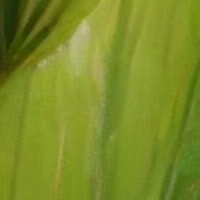}
        }
        \subfloat[][GT]
        {
            \includegraphics[width=.142\linewidth]{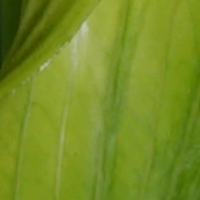}
        }
        

        \subfloat[][Short-focus]
        {
            \includegraphics[height=.142\linewidth, width=.213\linewidth]{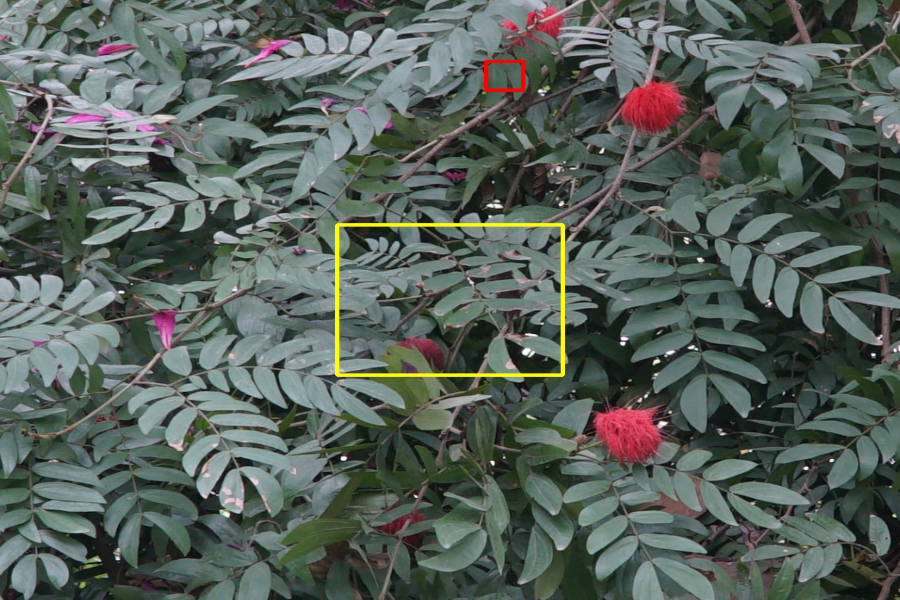}
        }
        \subfloat[][LR]
        {
            \includegraphics[width=.142\linewidth]{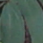}
        }
        \subfloat[][\cite{BSRGAN}]
        {
            \includegraphics[width=.142\linewidth]{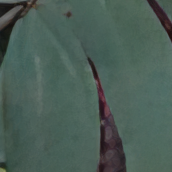}
        }
        \subfloat[][\cite{Real-ESRGAN}]
        {
            \includegraphics[width=.142\linewidth]{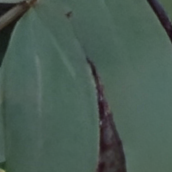}
        }
        \subfloat[][\cite{SRNTT}]
        {
            \includegraphics[width=.142\linewidth]{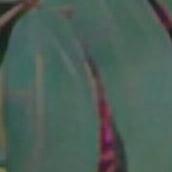}
        }
        \subfloat[][\cite{TTSR}]
        {
            \includegraphics[width=.142\linewidth]{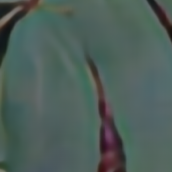}
        }
        
        \subfloat[][Telephoto]
        {
            \includegraphics[height=.142\linewidth, width=.213\linewidth]{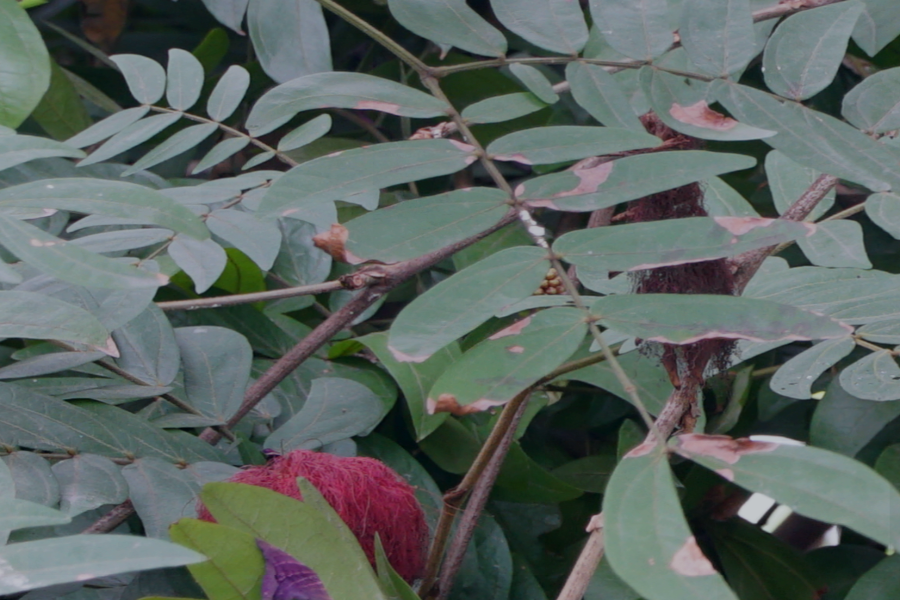}
        }
        \subfloat[][~\cite{C2-Matching}]
        {
            \includegraphics[width=.142\linewidth]{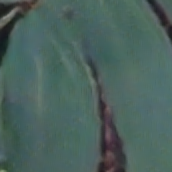}
        }
        \subfloat[][\cite{MASA-SR}]
        {
            \includegraphics[width=.142\linewidth]{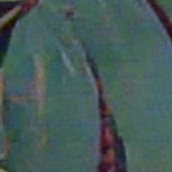}
        }
        \subfloat[][\cite{DCSR}]
        {
            \includegraphics[width=.142\linewidth]{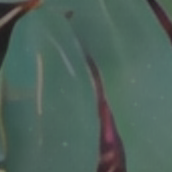}
        }
        \subfloat[][SelfDZSR]
        {
            \includegraphics[width=.142\linewidth]{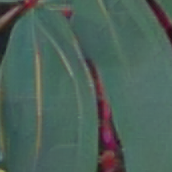}
        }
        \subfloat[][GT]
        {
            \includegraphics[width=.142\linewidth]{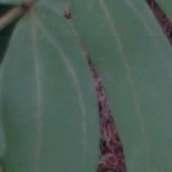}
        }
    \end{minipage}
    \caption{Visual comparison on \textbf{Sony} camera. In the short-focus image, the yellow box indicates the overlapped scene with the telephoto image, while the red box represents the selected LR patch. Our results in sub-figure (k) and (w) restore much more fine-scale edges.}
    \label{fig:Sony} 
\end{figure*}

  Tables~\ref{tab:Canon}$\sim$\ref{tab:Sony} show the quantitative results on four cameras, respectively.
  Our proposed model (SelfDZSR) achieves better results than most other methods, especially on \emph{Full-Image} and LPIPS metric.
  The visual comparison is carried out on the methods that are trained not only with $\ell_1$ (or $\ell_2$) loss.
  The comparisons on four cameras can be seen in Fig.~\ref{fig:Canon}$\sim$\ref{fig:Sony}, respectively.

\section{Additional Visual Comparison on Nikon Camera and CameraFusion Dataset} \label{sec:visual}
\begin{figure*}
    \begin{minipage}[ht]{\linewidth}
        \centering
        \subfloat[][Short-focus]
        {
            \includegraphics[height=.142\linewidth, width=.213\linewidth]{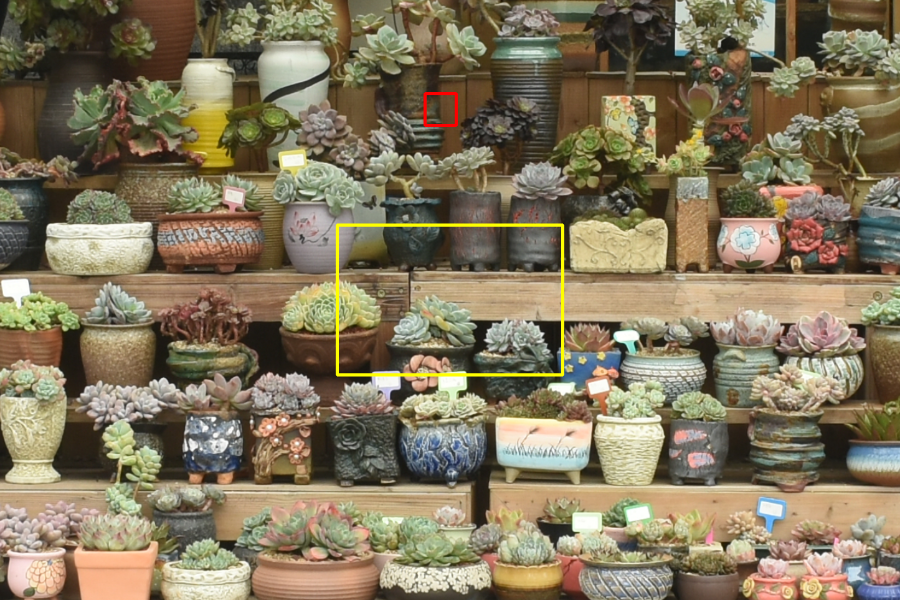}
        }
        \subfloat[][LR]
        {
            \includegraphics[width=.142\linewidth]{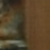}
        }
        \subfloat[][\cite{BSRGAN}]
        {
            \includegraphics[width=.142\linewidth]{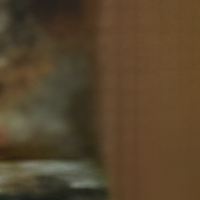}
        }
        \subfloat[][\cite{Real-ESRGAN}]
        {
            \includegraphics[width=.142\linewidth]{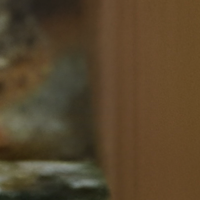}
        }
        \subfloat[][\cite{SRNTT}]
        {
            \includegraphics[width=.142\linewidth]{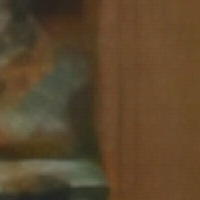}
        }
        \subfloat[][\cite{TTSR}]
        {
            \includegraphics[width=.142\linewidth]{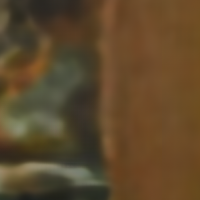}
        }
        
        \subfloat[][Telephoto]
        {
            \includegraphics[height=.142\linewidth, width=.213\linewidth]{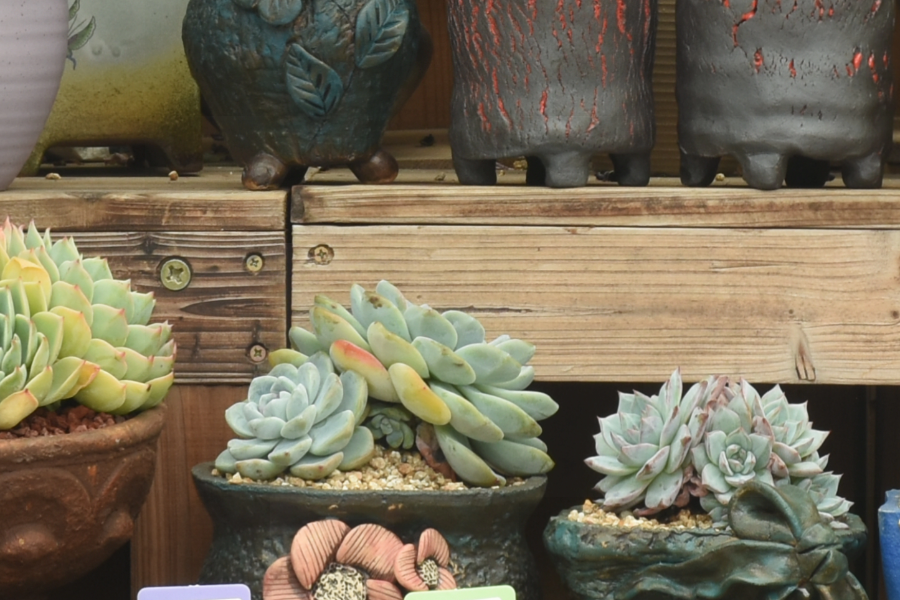}
        }
        \subfloat[][~\cite{C2-Matching}]
        {
            \includegraphics[width=.142\linewidth]{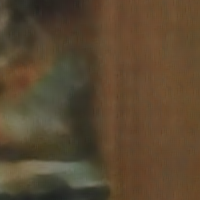}
        }
        \subfloat[][\cite{MASA-SR}]
        {
            \includegraphics[width=.142\linewidth]{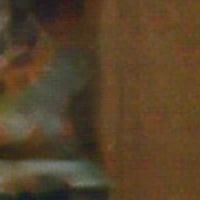}
        }
        \subfloat[][\cite{DCSR}]
        {
            \includegraphics[width=.142\linewidth]{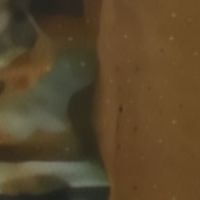}
        }
        \subfloat[][SelfDZSR]
        {
            \includegraphics[width=.142\linewidth]{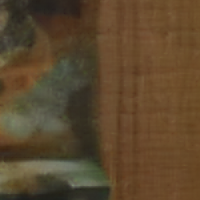}
        }
        \subfloat[][GT]
        {
            \includegraphics[width=.142\linewidth]{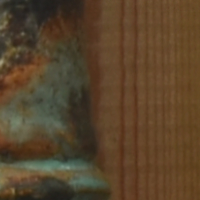}
        }
        

        \subfloat[][Short-focus]
        {
            \includegraphics[height=.142\linewidth, width=.213\linewidth]{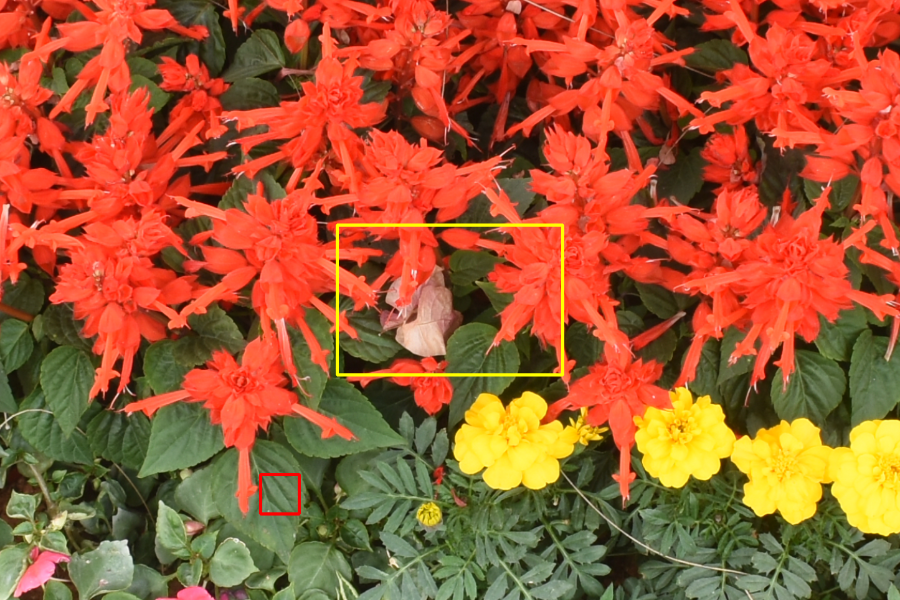}
        }
        \subfloat[][LR]
        {
            \includegraphics[width=.142\linewidth]{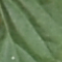}
        }
        \subfloat[][\cite{BSRGAN}]
        {
            \includegraphics[width=.142\linewidth]{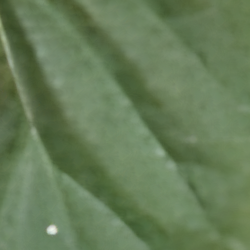}
        }
        \subfloat[][\cite{Real-ESRGAN}]
        {
            \includegraphics[width=.142\linewidth]{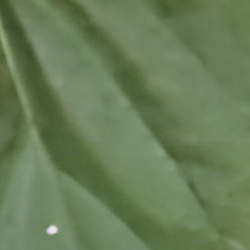}
        }
        \subfloat[][\cite{SRNTT}]
        {
            \includegraphics[width=.142\linewidth]{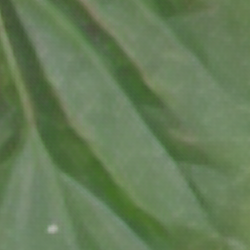}
        }
        \subfloat[][\cite{TTSR}]
        {
            \includegraphics[width=.142\linewidth]{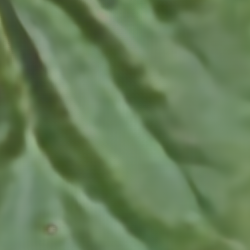}
        }
        
        \subfloat[][Telephoto]
        {
            \includegraphics[height=.142\linewidth, width=.213\linewidth]{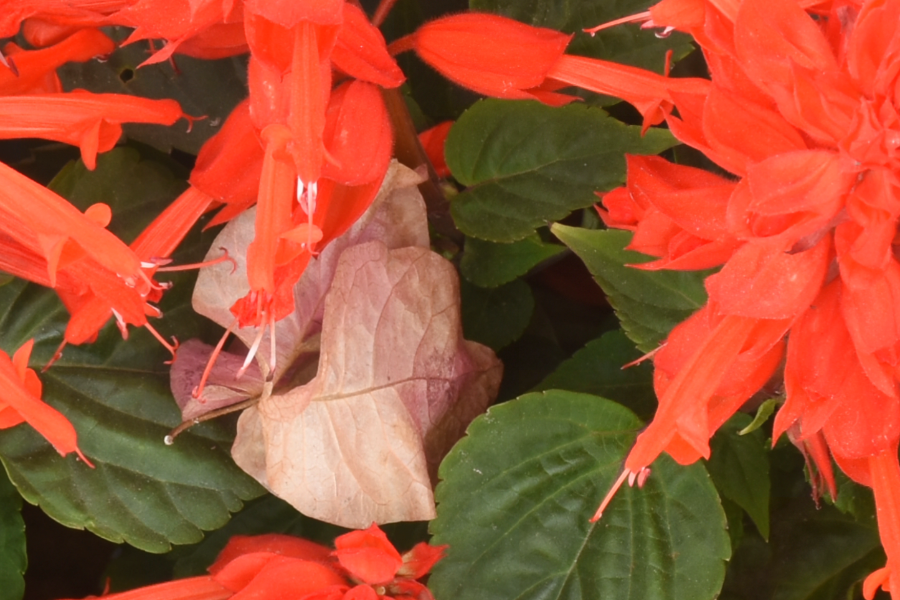}
        }
        \subfloat[][~\cite{C2-Matching}]
        {
            \includegraphics[width=.142\linewidth]{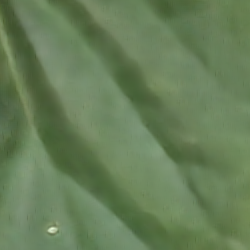}
        }
        \subfloat[][\cite{MASA-SR}]
        {
            \includegraphics[width=.142\linewidth]{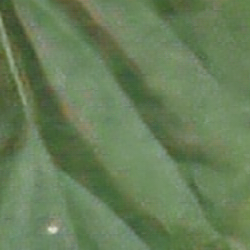}
        }
        \subfloat[][\cite{DCSR}]
        {
            \includegraphics[width=.142\linewidth]{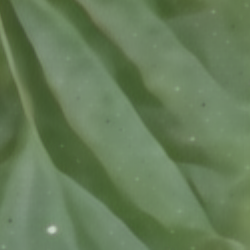}
        }
        \subfloat[][SelfDZSR]
        {
            \includegraphics[width=.142\linewidth]{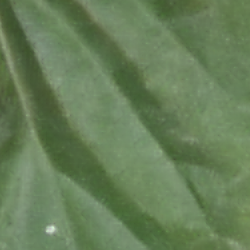}
        }
        \subfloat[][GT]
        {
            \includegraphics[width=.142\linewidth]{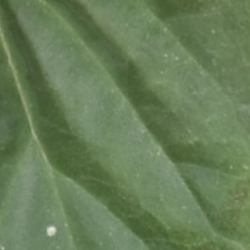}
        }
    \end{minipage}
    \caption{Visual comparison on \textbf{Nikon} camera. In the short-focus image, the yellow box indicates the overlapped scene with the telephoto image, while the red box represents the selected LR patch. Our result in sub-figure (k) restores much more details, and that in sub-figure (w) is more photo-realistic.}
    \label{fig:Nikon} 
\end{figure*}

\begin{figure*}
    \begin{minipage}[ht]{\linewidth}
        \centering
        \subfloat[][Short-focus]
        {
            \includegraphics[height=.142\linewidth, width=.213\linewidth]{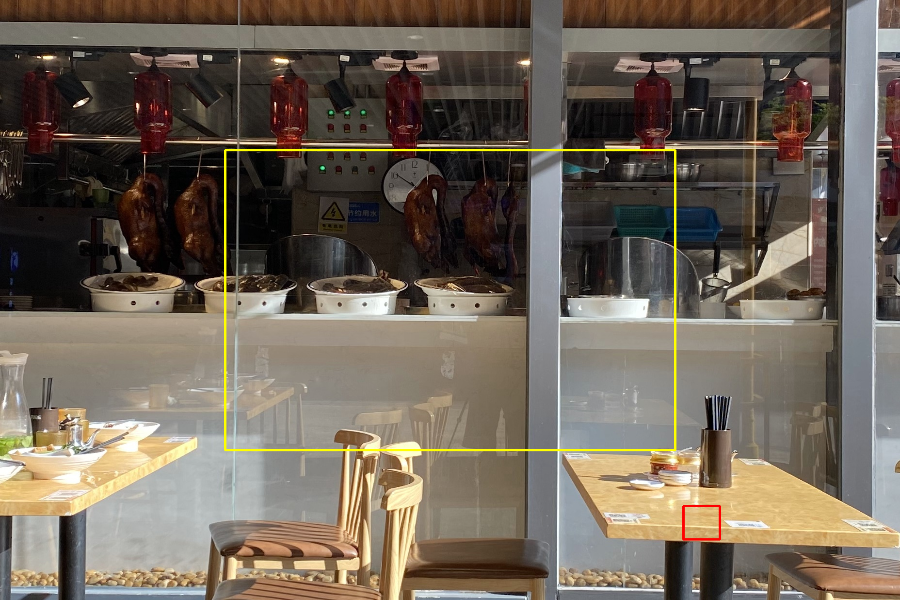}
        }
        \subfloat[][LR]
        {
            \includegraphics[width=.142\linewidth]{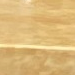}
        }
        \subfloat[][\cite{BSRGAN}]
        {
            \includegraphics[width=.142\linewidth]{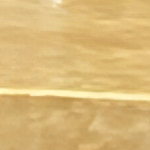}
        }
        \subfloat[][\cite{Real-ESRGAN}]
        {
            \includegraphics[width=.142\linewidth]{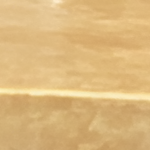}
        }
        \subfloat[][\cite{SRNTT}]
        {
            \includegraphics[width=.142\linewidth]{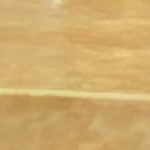}
        }
        \subfloat[][\cite{TTSR}]
        {
            \includegraphics[width=.142\linewidth]{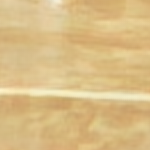}
        }
        
        \subfloat[][Telephoto]
        {
            \includegraphics[height=.142\linewidth, width=.213\linewidth]{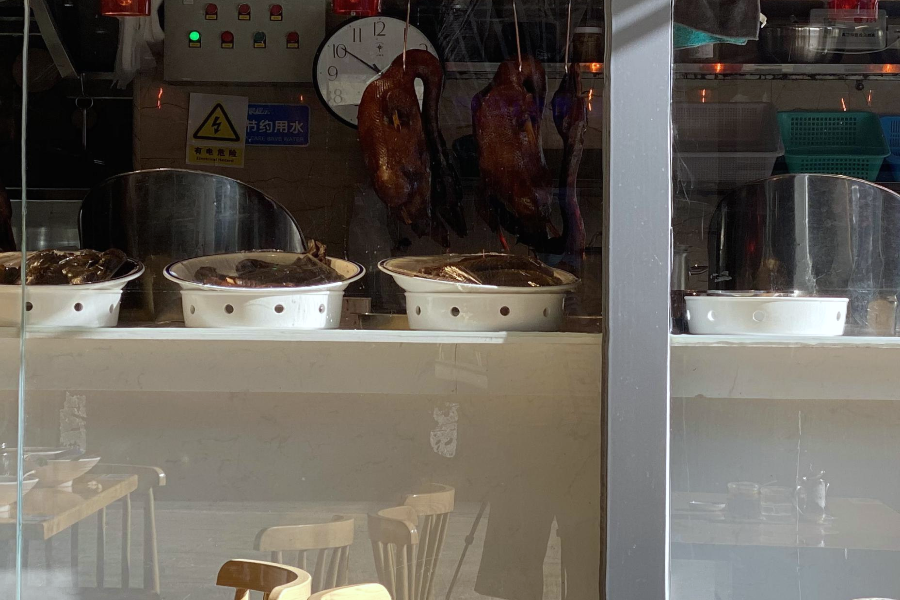}
        }
        \subfloat[][~\cite{C2-Matching}]
        {
            \includegraphics[width=.142\linewidth]{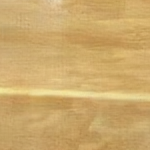}
        }
        \subfloat[][\cite{MASA-SR}]
        {
            \includegraphics[width=.142\linewidth]{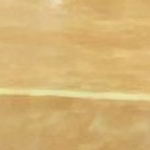}
        }
        \subfloat[][\cite{DCSR}]
        {
            \includegraphics[width=.142\linewidth]{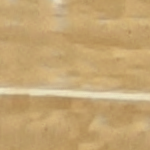}
        }
        \subfloat[][SelfDZSR]
        {
            \includegraphics[width=.142\linewidth]{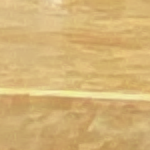}
        }
        \subfloat[][GT]
        {
            \includegraphics[width=.142\linewidth]{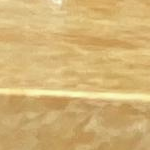}
        }
        
        
        \subfloat[][Short-focus]
        {
            \includegraphics[height=.142\linewidth, width=.213\linewidth]{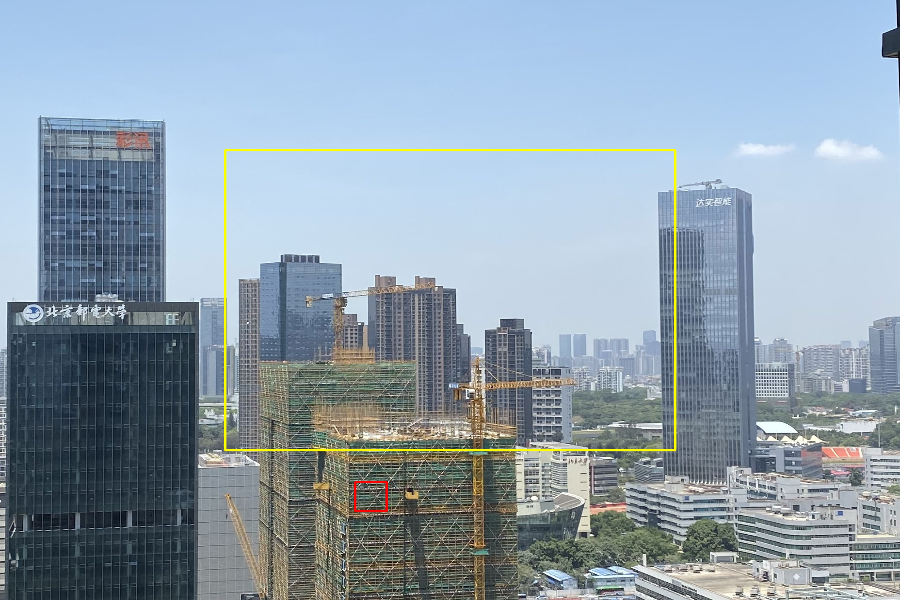}
        }
        \subfloat[][LR]
        {
            \includegraphics[width=.142\linewidth]{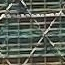}
        }
        \subfloat[][\cite{BSRGAN}]
        {
            \includegraphics[width=.142\linewidth]{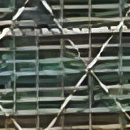}
        }
        \subfloat[][\cite{Real-ESRGAN}]
        {
            \includegraphics[width=.142\linewidth]{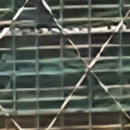}
        }
        \subfloat[][\cite{SRNTT}]
        {
            \includegraphics[width=.142\linewidth]{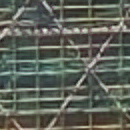}
        }
        \subfloat[][\cite{TTSR}]
        {
            \includegraphics[width=.142\linewidth]{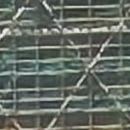}
        }
        
        \subfloat[][Telephoto]
        {
            \includegraphics[height=.142\linewidth, width=.213\linewidth]{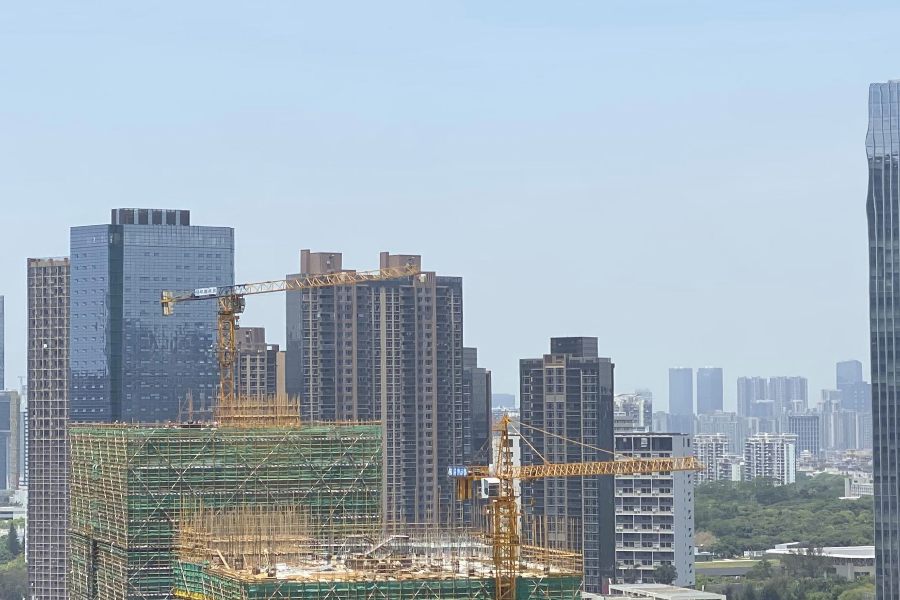}
        }
        \subfloat[][~\cite{C2-Matching}]
        {
            \includegraphics[width=.142\linewidth]{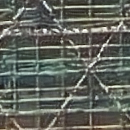}
        }
        \subfloat[][\cite{MASA-SR}]
        {
            \includegraphics[width=.142\linewidth]{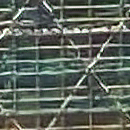}
        }
        \subfloat[][\cite{DCSR}]
        {
            \includegraphics[width=.142\linewidth]{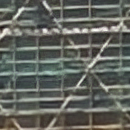}
        }
        \subfloat[][SelfDZSR]
        {
            \includegraphics[width=.142\linewidth]{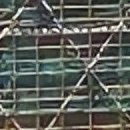}
        }
        \subfloat[][GT]
        {
            \includegraphics[width=.142\linewidth]{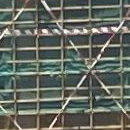}
        }        
    \end{minipage}
    \caption{Visual comparison on \textbf{CameraFusion} dataset. In the short-focus image, the yellow box indicates the overlapped scene with the telephoto image, while the red box represents the selected LR patch. Our result in sub-figure (k) restores much more textures, and that in sub-figure (w) is clearer and more photo-realistic.}
    \label{fig:cam} 
\end{figure*}

  The visual comparison is carried out on the methods that are trained not only with $\ell_1$ (or $\ell_2$) loss.
  In Fig.~\ref{fig:Nikon}, we show more qualitative comparison on Nikon camera images~\cite{CDC}.
  The visual comparison on CameraFusion dataset~\cite{DCSR} can be seen in Fig.~\ref{fig:cam}.
  The resolution of full LR images from the two test sets is 1$\sim$2K, so we select a patch for comparison. 

\clearpage

\begin{table*}[t] 
  \scriptsize
  \caption{Quantitative results on \textbf{Canon} camera with 17 images. Best results are highlighted by {\color{red}red}. The models trained only with $\ell_1$ (or $\ell_2$) loss are marked in gray. RefSR$^\dagger$ represents that the RefSR methods are trained in our self-supervised learning manner.}
  \label{tab:Canon}
  \centering\noindent
  \centering%
  \begin{center}
    \begin{tabular}{clccc}
      \toprule
      & Method
      & \tabincell{c}{\# Param\\(M)}
      & \tabincell{c}{\emph{Full-Image} \\ {{PSNR}{$\uparrow$} {/} {SSIM}{$\uparrow$} {/} {LPIPS}{$\downarrow$}}}
      & \tabincell{c}{\emph{Corner-Image} \\ {{PSNR}{$\uparrow$} {/} {SSIM}{$\uparrow$} {/} {LPIPS}{$\downarrow$}}} \\
        \hline
        \multirow{5}{*}{SISR}
        &  \cellcolor{Gray}EDSR~\cite{EDSR}  &  \cellcolor{Gray}43.1 &  \cellcolor{Gray}26.52 / 0.8399 / 0.342 &  \cellcolor{Gray}26.55 / 0.8383 / 0.342 \\ 
        &  \cellcolor{Gray}RCAN~\cite{RCAN}  &  \cellcolor{Gray}15.6 &  \cellcolor{Gray}26.69 / 0.8413 / 0.346 &  \cellcolor{Gray}26.73 / 0.8404 / 0.347 \\ 
        &  \cellcolor{Gray}CDC~\cite{CDC}    &  \cellcolor{Gray}39.9 &  \cellcolor{Gray}24.85 / 0.8378 / 0.384 &  \cellcolor{Gray}24.90 / 0.8366 / 0.385 \\ 
        & BSRGAN~\cite{BSRGAN}            & 16.7 & 25.39 / 0.8031 / 0.268 & 25.43 / 0.8017 / 0.268 \\ 
        & Real-ESRGAN~\cite{Real-ESRGAN}  & 16.7 & 24.64 / 0.8010 / 0.275 & 24.68 / 0.7988 / 0.276 \\ 
        \hline
        \multirow{10}{*}{RefSR$^\dagger$}  
        &  \cellcolor{Gray}SRNTT-$\ell_2$~\cite{SRNTT}   & \cellcolor{Gray}5.5    &  \cellcolor{Gray}26.22 / 0.8390 / 0.348 &  \cellcolor{Gray}26.24 / 0.8378 / 0.351 \\ 
        & SRNTT~\cite{SRNTT}                 & 5.5    & 26.25 / 0.8268 / 0.295 & 26.28 / 0.8258 / 0.293  \\ 
        &  \cellcolor{Gray}TTSR-$\ell_1$~\cite{TTSR}     &  \cellcolor{Gray}7.3 &  \cellcolor{Gray}23.97 / 0.8280 / 0.374 &  \cellcolor{Gray}23.93 / 0.8259 / 0.375 \\ 
        & TTSR~\cite{TTSR}                   & 7.3 & 23.75 / 0.7719 / 0.340 & 23.68 / 0.7695 / 0.338 \\ 
        &  \cellcolor{Gray}$C^2$-Matching-$\ell_1$~\cite{C2-Matching}  &  \cellcolor{Gray}8.9 &  \cellcolor{Gray}25.64 / 0.8383 / 0.357 &  \cellcolor{Gray}25.60 / 0.8373 / 0.358 \\ 
        & $C^2$-Matching~\cite{C2-Matching}  & 8.9 & 24.75 / 0.8180 / 0.329 & 24.72 / 0.8171 / 0.328 \\
        &  \cellcolor{Gray}MASA-$\ell_1$~\cite{MASA-SR}  &  \cellcolor{Gray}4.0 &  \cellcolor{Gray}26.54 / 0.8398 / 0.338 &  \cellcolor{Gray}26.58 / 0.8390 / 0.339 \\ 
        & MASA~\cite{MASA-SR}                & 4.0 & 27.19 / 0.8006 / 0.306  & 27.24 / 0.7994 / 0.305 \\
        &  \cellcolor{Gray}DCSR-$\ell_1$~\cite{DCSR}    &  \cellcolor{Gray}3.2 &  \cellcolor{Gray}27.55 / 0.8363 / 0.338 &  \cellcolor{Gray}27.54 / 0.8377 / 0.330 \\
        & DCSR~\cite{DCSR}                   & 3.2  & 26.80 / 0.8265 / 0.275 & 26.79 / 0.8275 / 0.268 \\
        \hline
        \multirow{2}{*}{Ours}  
        &  \cellcolor{Gray}SelfDZSR-$\ell_1$  &  \cellcolor{Gray}3.2  & \cellcolor{Gray}{\color{red}28.13} / \cellcolor{Gray}{\color{red}0.8576} / \cellcolor{Gray}0.300 & \cellcolor{Gray}{\color{red}27.87} / \cellcolor{Gray}{\color{red}0.8465} / \cellcolor{Gray}0.321 \\ 
        & SelfDZSR   & 3.2  & 27.85 / 0.8386 / {\color{red}0.240} & 27.60 / 0.8274 / {\color{red}0.253} \\ 
      \bottomrule
    \end{tabular}
    \end{center}
\end{table*}

\begin{table*}[t] 
  \scriptsize
  \caption{Quantitative results on \textbf{Olympus} camera with 19 images. Best results are highlighted by {\color{red}red}. The models trained only with $\ell_1$ (or $\ell_2$) loss are marked in gray. RefSR$^\dagger$ represents that the RefSR methods are trained in our self-supervised learning manner.}
  \label{tab:Oly}
  \centering\noindent
  \centering%
  \begin{center}
    \begin{tabular}{clccc}
      \toprule
      & Method
      & \tabincell{c}{\# Param\\(M)}
      & \tabincell{c}{\emph{Full-Image} \\ {{PSNR}{$\uparrow$} {/} {SSIM}{$\uparrow$} {/} {LPIPS}{$\downarrow$}}}
      & \tabincell{c}{\emph{Corner-Image} \\ {{PSNR}{$\uparrow$} {/} {SSIM}{$\uparrow$} {/} {LPIPS}{$\downarrow$}}} \\
        \hline
        \multirow{5}{*}{SISR}
        &  \cellcolor{Gray}EDSR~\cite{EDSR}  &  \cellcolor{Gray}43.1 &  \cellcolor{Gray}26.99 / 0.7960 / 0.452 &  \cellcolor{Gray}26.99 / 0.7917 / 0.451 \\ 
        &  \cellcolor{Gray}RCAN~\cite{RCAN}  &  \cellcolor{Gray}15.6 &  \cellcolor{Gray}27.54 / 0.8038 / 0.452 &  \cellcolor{Gray}27.54 / 0.7995 / 0.451 \\ 
        &  \cellcolor{Gray}CDC~\cite{CDC}    &  \cellcolor{Gray}39.9 &  \cellcolor{Gray}27.31 / 0.8030 / 0.466 &  \cellcolor{Gray}27.30 / 0.7988 / 0.467 \\ 
        & BSRGAN~\cite{BSRGAN}            & 16.7 & 25.76 / 0.7422 / 0.341 & 25.75 / 0.7388 / 0.341 \\ 
        & Real-ESRGAN~\cite{Real-ESRGAN}  & 16.7 & 26.00 / 0.7545 / 0.323 & 25.98 / 0.7517 / 0.321 \\ 
        \hline
        \multirow{10}{*}{RefSR$^\dagger$}  
        &  \cellcolor{Gray}SRNTT-$\ell_2$~\cite{SRNTT}   & \cellcolor{Gray}5.5    &  \cellcolor{Gray}26.51 / 0.7928 / 0.442 &  \cellcolor{Gray}26.49 / 0.7879 / 0.441 \\  
        & SRNTT~\cite{SRNTT}                 & 5.5    & 27.04 / 0.7870 / 0.357 & 27.03 / 0.7823 / 0.354  \\ 
        &  \cellcolor{Gray}TTSR-$\ell_1$~\cite{TTSR}     &  \cellcolor{Gray}7.3 &  \cellcolor{Gray}25.44 / 0.7790 / 0.469 &  \cellcolor{Gray}25.39 / 0.7756 / 0.466 \\ 
        & TTSR~\cite{TTSR}                   & 7.3 & 25.12 / 0.7736 / 0.377 & 25.08 / 0.7693 / 0.370 \\ 
        &  \cellcolor{Gray}$C^2$-Matching-$\ell_1$~\cite{C2-Matching}  &  \cellcolor{Gray}8.9 &  \cellcolor{Gray}26.65 / 0.8001 / 0.448 &  \cellcolor{Gray}26.62 / 0.7960 / 0.446 \\ 
        & $C^2$-Matching~\cite{C2-Matching}  & 8.9 & 26.51 / 0.7728 / 0.380 & 26.49 / 0.7682 / 0.378 \\
        &  \cellcolor{Gray}MASA-$\ell_1$~\cite{MASA-SR}  &  \cellcolor{Gray}4.0 &  \cellcolor{Gray}27.17 / 0.7982 / 0.423 &  \cellcolor{Gray}27.17 / 0.7937 / 0.423 \\ 
        & MASA~\cite{MASA-SR}                & 4.0 & 26.66 / 0.7393 / 0.351  & 26.66 / 0.7348 / 0.350 \\
        &  \cellcolor{Gray}DCSR-$\ell_1$~\cite{DCSR}    &  \cellcolor{Gray}3.2 &  \cellcolor{Gray}27.74 / 0.7987 / 0.437 &  \cellcolor{Gray}{\color{red}27.73} / 0.7973 / 0.426 \\
        & DCSR~\cite{DCSR}                   & 3.2  & 27.41 / 0.7894 / 0.351 & 27.39 / 0.7882 / 0.342 \\
        \hline
        \multirow{2}{*}{Ours}  
        &  \cellcolor{Gray}SelfDZSR-$\ell_1$  &  \cellcolor{Gray}3.2  & \cellcolor{Gray}{\color{red}27.79} / \cellcolor{Gray}{\color{red}0.8124} / \cellcolor{Gray}0.395 & \cellcolor{Gray}27.56 / \cellcolor{Gray}{\color{red}0.7980} / \cellcolor{Gray}0.420 \\ 
        & SelfDZSR   & 3.2  & 27.34 / 0.7862 / {\color{red}0.292} & 27.12 / 0.7722 / {\color{red}0.310} \\ 
      \bottomrule
    \end{tabular}
    \end{center}
\end{table*}

\begin{table*}[t] 
  \scriptsize
  \caption{Quantitative results on \textbf{Panasonic} camera with 20 images.  Best results are highlighted by {\color{red}red}. The models trained only with $\ell_1$ (or $\ell_2$) loss are marked in gray. RefSR$^\dagger$ represents that the RefSR methods are trained in our self-supervised learning manner.}
  \label{tab:Pan}
  \centering\noindent
  \centering%
  \begin{center}
    \begin{tabular}{clccc}
      \toprule
      & Method
      & \tabincell{c}{\# Param\\(M)}
      & \tabincell{c}{\emph{Full-Image} \\ {{PSNR}{$\uparrow$} {/} {SSIM}{$\uparrow$} {/} {LPIPS}{$\downarrow$}}}
      & \tabincell{c}{\emph{Corner-Image} \\ {{PSNR}{$\uparrow$} {/} {SSIM}{$\uparrow$} {/} {LPIPS}{$\downarrow$}}} \\
        \hline
        \multirow{5}{*}{SISR}
        &  \cellcolor{Gray}EDSR~\cite{EDSR}  &  \cellcolor{Gray}43.1 &  \cellcolor{Gray}27.04 / 0.7994 / 0.379 &  \cellcolor{Gray}27.15 / 0.7964 / 0.380 \\ 
        &  \cellcolor{Gray}RCAN~\cite{RCAN}  &  \cellcolor{Gray}15.6 &  \cellcolor{Gray}27.26 / 0.8055 / 0.381 &  \cellcolor{Gray}27.36 / 0.8027 / 0.383 \\ 
        &  \cellcolor{Gray}CDC~\cite{CDC}    &  \cellcolor{Gray}39.9 &  \cellcolor{Gray}27.02 / 0.7981 / 0.414 &  \cellcolor{Gray}27.12 / 0.7949 / 0.414 \\ 
        & BSRGAN~\cite{BSRGAN}            & 16.7 & 26.27 / 0.7520 / 0.288 & 26.40 / 0.7482 / 0.288 \\ 
        & Real-ESRGAN~\cite{Real-ESRGAN}  & 16.7 & 26.20 / 0.7625 / 0.275 & 26.31 / 0.7592 / 0.274 \\ 
        \hline
        \multirow{10}{*}{RefSR$^\dagger$}  
        &  \cellcolor{Gray}SRNTT-$\ell_2$~\cite{SRNTT}   & \cellcolor{Gray}5.5    &  \cellcolor{Gray}27.08 / 0.7988 / 0.374 &  \cellcolor{Gray}27.18 / 0.7960 / 0.375 \\ 
        & SRNTT~\cite{SRNTT}                 & 5.5    & 27.14 / 0.7862 / 0.307 & 27.22 / 0.7829 / 0.306  \\ 
        &  \cellcolor{Gray}TTSR-$\ell_1$~\cite{TTSR}     &  \cellcolor{Gray}7.3 &  \cellcolor{Gray}26.21 / 0.7859 / 0.383 &  \cellcolor{Gray}26.26 / 0.7842 / 0.385 \\ 
        & TTSR~\cite{TTSR}                   & 7.3 & 25.24 / 0.7558 / 0.329 & 25.26 / 0.7537 / 0.326 \\ 
        &  \cellcolor{Gray}$C^2$-Matching-$\ell_1$~\cite{C2-Matching}  &  \cellcolor{Gray}8.9 &  \cellcolor{Gray}26.61 / 0.8032 / 0.378 &  \cellcolor{Gray}26.71 / 0.7994 / 0.381 \\ 
        & $C^2$-Matching~\cite{C2-Matching}  & 8.9 & 25.70 / 0.7649 / 0.340 & 25.78 / 0.7596 / 0.342 \\
        &  \cellcolor{Gray}MASA-$\ell_1$~\cite{MASA-SR}  &  \cellcolor{Gray}4.0 &  \cellcolor{Gray}26.94 / 0.7997 / 0.363 &  \cellcolor{Gray}27.00 / 0.7958 / 0.365 \\ 
        & MASA~\cite{MASA-SR}                & 4.0 & 26.93 / 0.7388 / 0.299  & 27.04 / 0.7365 / 0.299 \\
        &  \cellcolor{Gray}DCSR-$\ell_1$~\cite{DCSR}    &  \cellcolor{Gray}3.2 &  \cellcolor{Gray}26.58 / 0.7640 / 0.398 &  \cellcolor{Gray}26.54 / 0.7632 / 0.390 \\
        & DCSR~\cite{DCSR}                   & 3.2  & 26.40 / 0.7543 / 0.315 & 26.36 / 0.7528 / 0.308 \\
        \hline
        \multirow{2}{*}{Ours}  
        &  \cellcolor{Gray}SelfDZSR-$\ell_1$  &  \cellcolor{Gray}3.2  &  \cellcolor{Gray}{\color{red}27.90} / {\color{red}0.8164} / 0.337 &  \cellcolor{Gray}{\color{red}27.67} / {\color{red}0.8001} / 0.361 \\ 
        & SelfDZSR   & 3.2  & 27.41 / 0.7836 / {\color{red}0.250} & 27.21 / 0.7674 / {\color{red}0.265} \\ 
      \bottomrule
    \end{tabular}
    \end{center}
\end{table*}

\begin{table*}[t] 
  \scriptsize
  \caption{Quantitative results on \textbf{Sony} camera with 17 images.  Best results are highlighted by {\color{red}red}. The models trained only with $\ell_1$ (or $\ell_2$) loss are marked in gray. RefSR$^\dagger$ represents that the RefSR methods are trained in our self-supervised learning manner.}
  \label{tab:Sony}
  \centering\noindent
  \centering%
  \begin{center}
    \begin{tabular}{clccc}
      \toprule
      & Method
      & \tabincell{c}{\# Param\\(M)}
      & \tabincell{c}{\emph{Full-Image} \\ {{PSNR}{$\uparrow$} {/} {SSIM}{$\uparrow$} {/} {LPIPS}{$\downarrow$}}}
      & \tabincell{c}{\emph{Corner-Image} \\ {{PSNR}{$\uparrow$} {/} {SSIM}{$\uparrow$} {/} {LPIPS}{$\downarrow$}}} \\
        \hline
        \multirow{5}{*}{SISR}
        &  \cellcolor{Gray}EDSR~\cite{EDSR}  &  \cellcolor{Gray}43.1 &  \cellcolor{Gray}27.12 / 0.8173 / 0.337 &  \cellcolor{Gray}27.13 / 0.8195 / 0.331 \\ 
        &  \cellcolor{Gray}RCAN~\cite{RCAN}  &  \cellcolor{Gray}15.6 &  \cellcolor{Gray}27.42 / 0.8248 / 0.333 &  \cellcolor{Gray}27.40 / 0.8274 / 0.326 \\ 
        &  \cellcolor{Gray}CDC~\cite{CDC}    &  \cellcolor{Gray}39.9 &  \cellcolor{Gray}27.27 / 0.8207 / 0.357 &  \cellcolor{Gray}27.27 / 0.8228 / 0.351 \\ 
        & BSRGAN~\cite{BSRGAN}            & 16.7 & 26.58 / 0.7732 / 0.284 & 26.57 / 0.7775 / 0.279 \\ 
        & Real-ESRGAN~\cite{Real-ESRGAN}  & 16.7 & 26.20 / 0.7816 / 0.262 & 26.18 / 0.7876 / 0.256 \\ 
        \hline
        \multirow{10}{*}{RefSR$^\dagger$}  
        &  \cellcolor{Gray}SRNTT-$\ell_2$~\cite{SRNTT}   & \cellcolor{Gray}5.5    &  \cellcolor{Gray}26.20 / 0.8103 / 0.337 &  \cellcolor{Gray}26.18 / 0.8138 / 0.331 \\ 
        & SRNTT~\cite{SRNTT}                 & 5.5    & 26.24 / 0.7969 / 0.290 & 26.23 / 0.8001 / 0.283  \\ 
        &  \cellcolor{Gray}TTSR-$\ell_1$~\cite{TTSR}     &  \cellcolor{Gray}7.3 &  \cellcolor{Gray}25.86 / 0.8152 / 0.333 &  \cellcolor{Gray}25.82 / 0.8195 / 0.327 \\ 
        & TTSR~\cite{TTSR}                   & 7.3 & 24.91 / 0.7326 / 0.315 & 24.86 / 0.7353 / 0.310 \\ 
        &  \cellcolor{Gray}$C^2$-Matching-$\ell_1$~\cite{C2-Matching}  &  \cellcolor{Gray}8.9 &  \cellcolor{Gray}26.78 / 0.8221 / 0.327 &  \cellcolor{Gray}26.73 / 0.8254 / 0.322 \\ 
        & $C^2$-Matching~\cite{C2-Matching}  & 8.9 & 26.49 / 0.7813 / 0.298 & 26.44 / 0.7848 / 0.289 \\
        &  \cellcolor{Gray}MASA-$\ell_1$~\cite{MASA-SR}  &  \cellcolor{Gray}4.0 &  \cellcolor{Gray}27.06 / 0.8149 / 0.306 &  \cellcolor{Gray}27.06 / 0.8189 / 0.301 \\ 
        & MASA~\cite{MASA-SR}                & 4.0 & 25.85 / 0.7075 / 0.325  & 25.84 / 0.7106 / 0.318 \\
        &  \cellcolor{Gray}DCSR-$\ell_1$~\cite{DCSR}    &  \cellcolor{Gray}3.2 &  \cellcolor{Gray}28.49 / 0.8216 / 0.335 &  \cellcolor{Gray}{\color{red}28.45} / 0.8237 / 0.330 \\
        & DCSR~\cite{DCSR}                   & 3.2  & 28.08 / 0.8128 / 0.272 & 28.03 / 0.8147 / 0.269 \\
        \hline
        \multirow{2}{*}{Ours}  
        &  \cellcolor{Gray}SelfDZSR-$\ell_1$  &  \cellcolor{Gray}3.2  &  \cellcolor{Gray}{\color{red}28.22} / {\color{red}0.8311} / 0.292 &  \cellcolor{Gray}28.34 / {\color{red}0.8359} / 0.303 \\ 
        & SelfDZSR   & 3.2  & 27.41 / 0.7921 / {\color{red}0.246} & 27.47 / 0.7948 / {\color{red}0.252} \\ 
      \bottomrule
    \end{tabular}
    \end{center}
\end{table*}

\begin{figure*}
    \begin{minipage}[ht]{\linewidth}
        \centering
        \subfloat[][Short-focus]
        {
            \includegraphics[height=.142\linewidth, width=.213\linewidth]{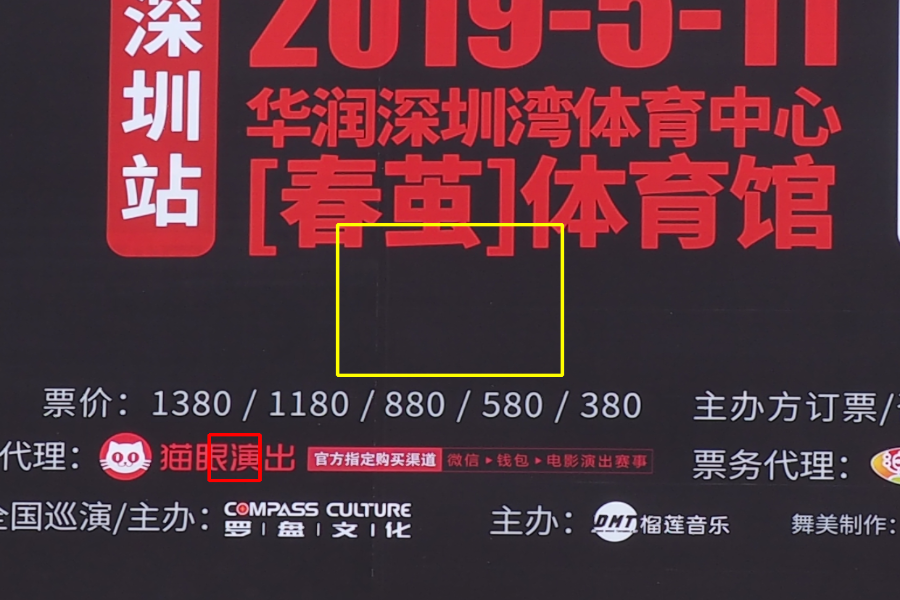}
        }
        \subfloat[][LR]
        {
            \includegraphics[width=.142\linewidth]{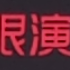}
        }
        \subfloat[][\cite{BSRGAN}]
        {
            \includegraphics[width=.142\linewidth]{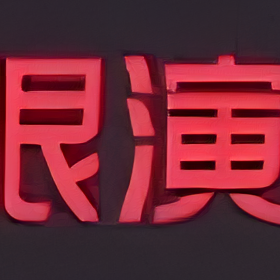}
        }
        \subfloat[][\cite{Real-ESRGAN}]
        {
            \includegraphics[width=.142\linewidth]{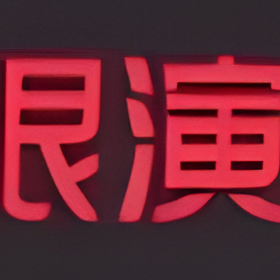}
        }
        \subfloat[][\cite{SRNTT}]
        {
            \includegraphics[width=.142\linewidth]{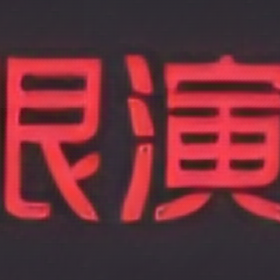}
        }
        \subfloat[][\cite{TTSR}]
        {
            \includegraphics[width=.142\linewidth]{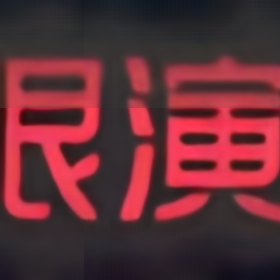}
        }
        
        \subfloat[][Telephoto]
        {
            \includegraphics[height=.142\linewidth, width=.213\linewidth]{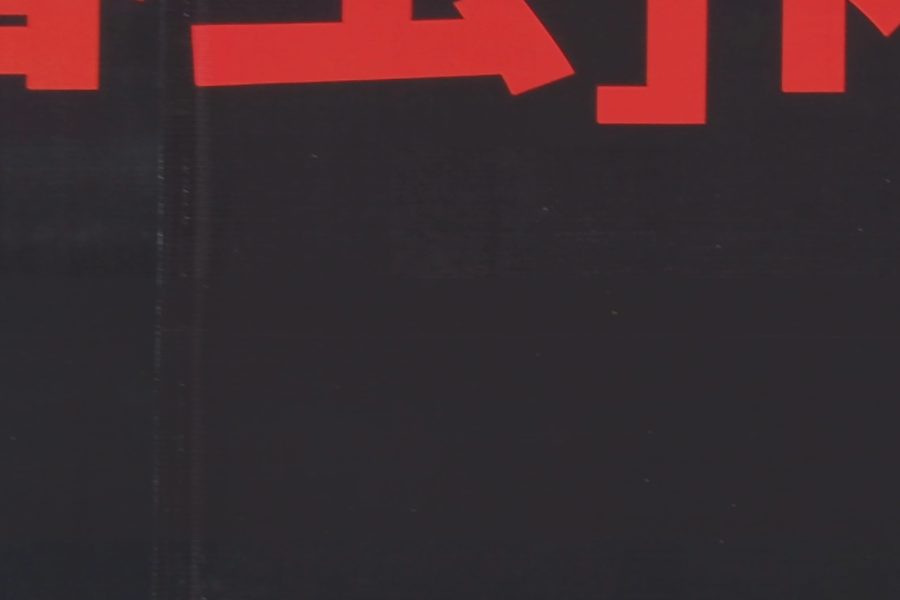}
        }
        \subfloat[][~\cite{C2-Matching}]
        {
            \includegraphics[width=.142\linewidth]{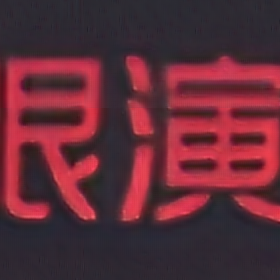}
        }
        \subfloat[][\cite{MASA-SR}]
        {
            \includegraphics[width=.142\linewidth]{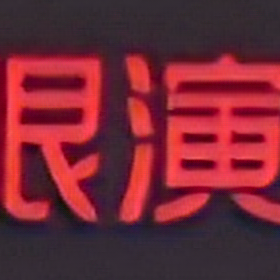}
        }
        \subfloat[][\cite{DCSR}]
        {
            \includegraphics[width=.142\linewidth]{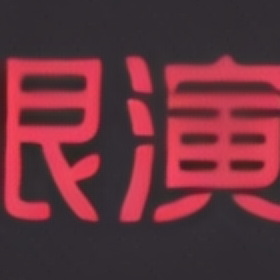}
        }
        \subfloat[][SelfDZSR]
        {
            \includegraphics[width=.142\linewidth]{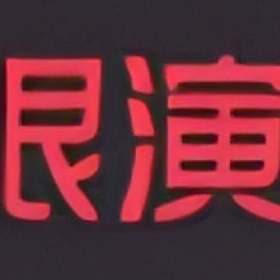}
        }
        \subfloat[][GT]
        {
            \includegraphics[width=.142\linewidth]{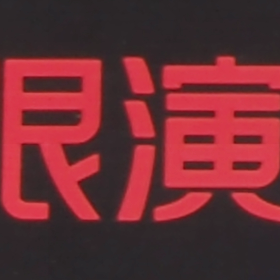}
        }
        

        \subfloat[][Short-focus]
        {
            \includegraphics[height=.142\linewidth, width=.213\linewidth]{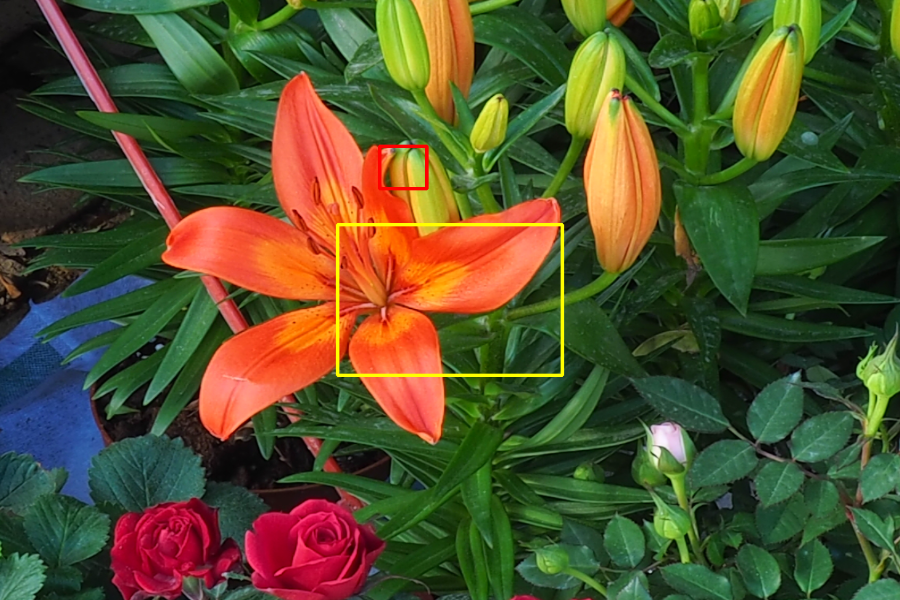}
        }
        \subfloat[][LR]
        {
            \includegraphics[width=.142\linewidth]{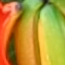}
        }
        \subfloat[][\cite{BSRGAN}]
        {
            \includegraphics[width=.142\linewidth]{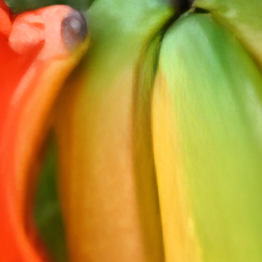}
        }
        \subfloat[][\cite{Real-ESRGAN}]
        {
            \includegraphics[width=.142\linewidth]{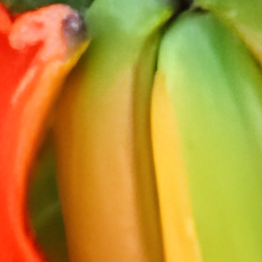}
        }
        \subfloat[][\cite{SRNTT}]
        {
            \includegraphics[width=.142\linewidth]{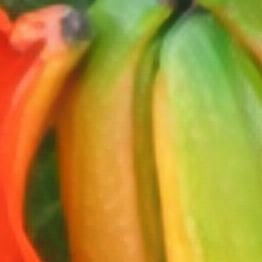}
        }
        \subfloat[][\cite{TTSR}]
        {
            \includegraphics[width=.142\linewidth]{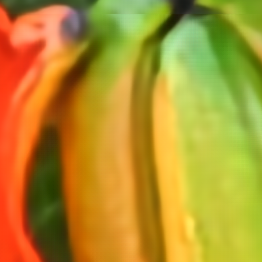}
        }
        
        \subfloat[][Telephoto]
        {
            \includegraphics[height=.142\linewidth, width=.213\linewidth]{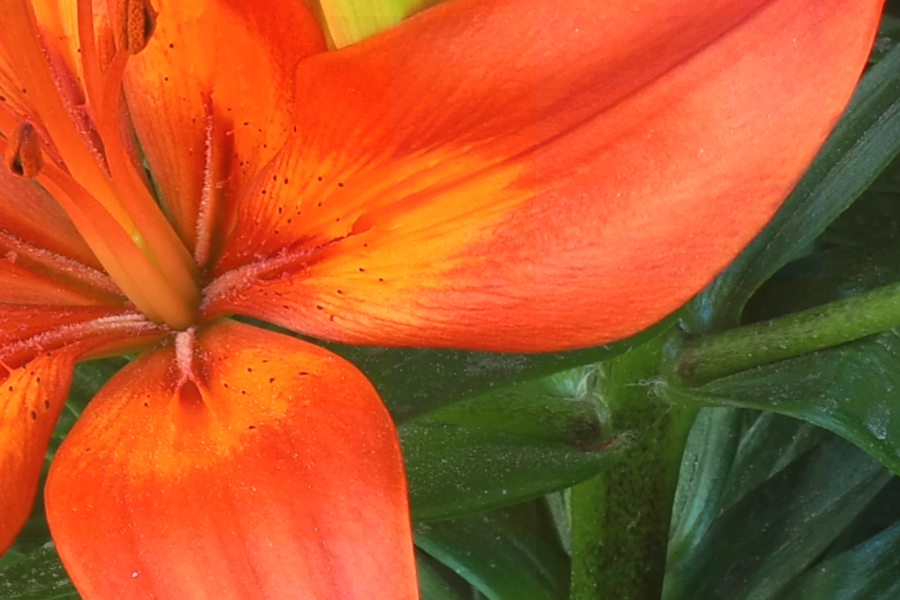}
        }
        \subfloat[][~\cite{C2-Matching}]
        {
            \includegraphics[width=.142\linewidth]{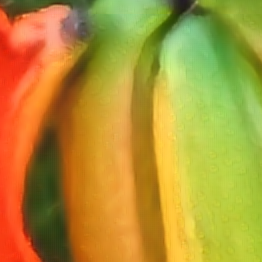}
        }
        \subfloat[][\cite{MASA-SR}]
        {
            \includegraphics[width=.142\linewidth]{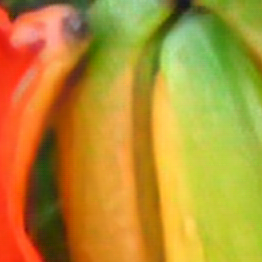}
        }
        \subfloat[][\cite{DCSR}]
        {
            \includegraphics[width=.142\linewidth]{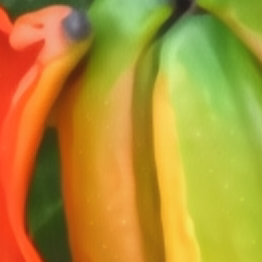}
        }
        \subfloat[][SelfDZSR]
        {
            \includegraphics[width=.142\linewidth]{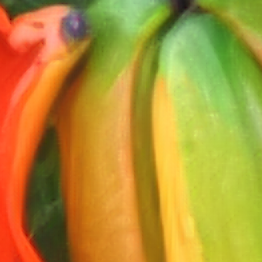}
        }
        \subfloat[][GT]
        {
            \includegraphics[width=.142\linewidth]{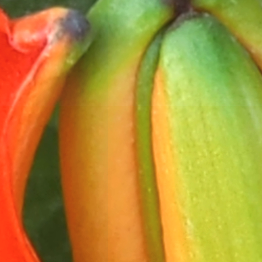}
        }
    \end{minipage}
    \caption{Visual comparison on \textbf{Olympus} camera. In the short-focus image, the yellow box indicates the overlapped scene with the telephoto image, while the red box represents the selected LR patch. Our result in sub-figure (k) is clearer, and that in sub-figure (w) is more photo-realistic.}
    \label{fig:Oly} 
\end{figure*}

\begin{figure*}
    \begin{minipage}[ht]{\linewidth}
        \centering
        \subfloat[][Short-focus]
        {
            \includegraphics[height=.142\linewidth, width=.213\linewidth]{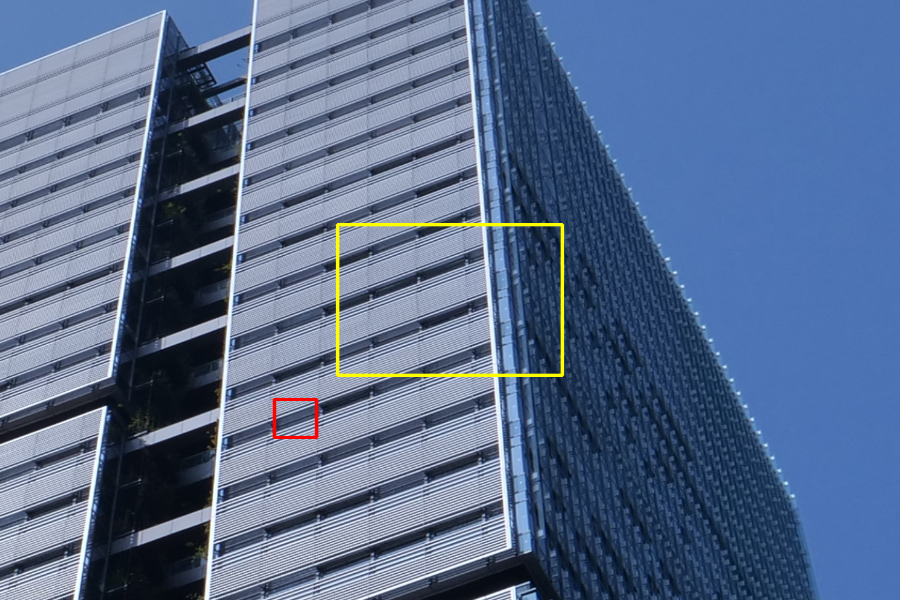}
        }
        \subfloat[][LR]
        {
            \includegraphics[width=.142\linewidth]{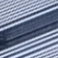}
        }
        \subfloat[][\cite{BSRGAN}]
        {
            \includegraphics[width=.142\linewidth]{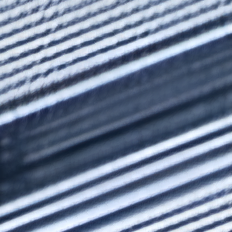}
        }
        \subfloat[][\cite{Real-ESRGAN}]
        {
            \includegraphics[width=.142\linewidth]{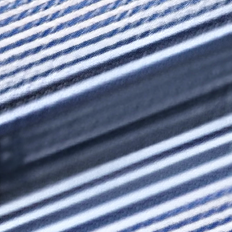}
        }
        \subfloat[][\cite{SRNTT}]
        {
            \includegraphics[width=.142\linewidth]{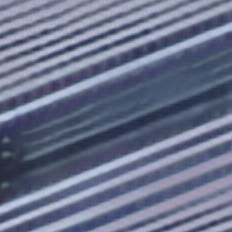}
        }
        \subfloat[][\cite{TTSR}]
        {
            \includegraphics[width=.142\linewidth]{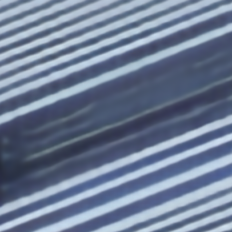}
        }
        
        \subfloat[][Telephoto]
        {
            \includegraphics[height=.142\linewidth, width=.213\linewidth]{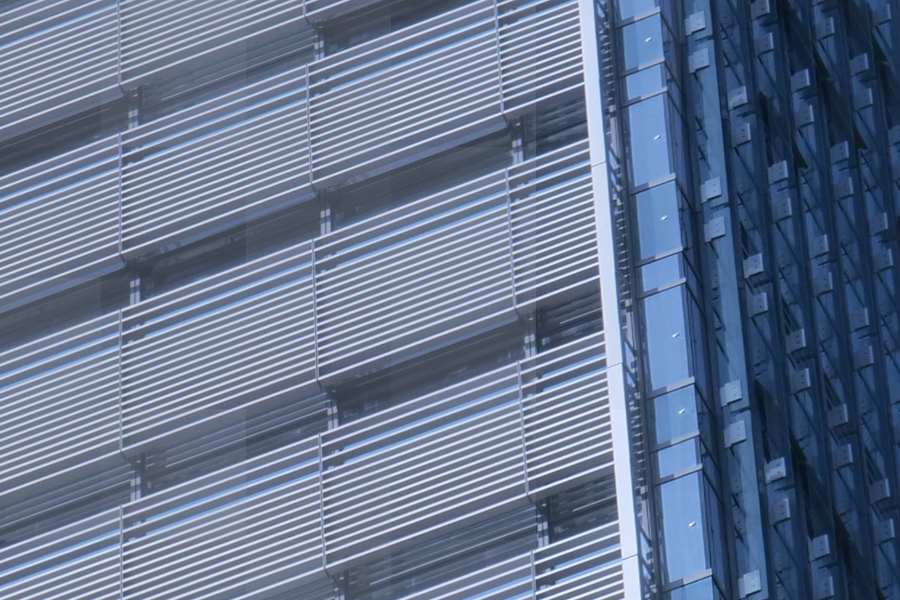}
        }
        \subfloat[][~\cite{C2-Matching}]
        {
            \includegraphics[width=.142\linewidth]{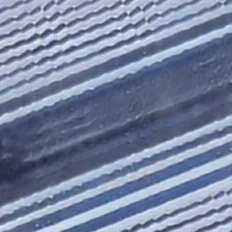}
        }
        \subfloat[][\cite{MASA-SR}]
        {
            \includegraphics[width=.142\linewidth]{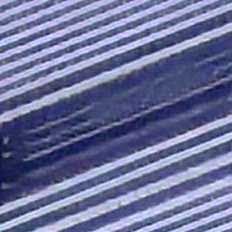}
        }
        \subfloat[][\cite{DCSR}]
        {
            \includegraphics[width=.142\linewidth]{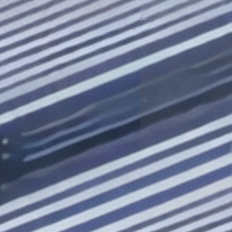}
        }
        \subfloat[][SelfDZSR]
        {
            \includegraphics[width=.142\linewidth]{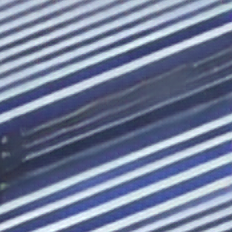}
        }
        \subfloat[][GT]
        {
            \includegraphics[width=.142\linewidth]{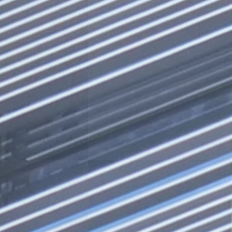}
        }
        

        \subfloat[][Short-focus]
        {
            \includegraphics[height=.142\linewidth, width=.213\linewidth]{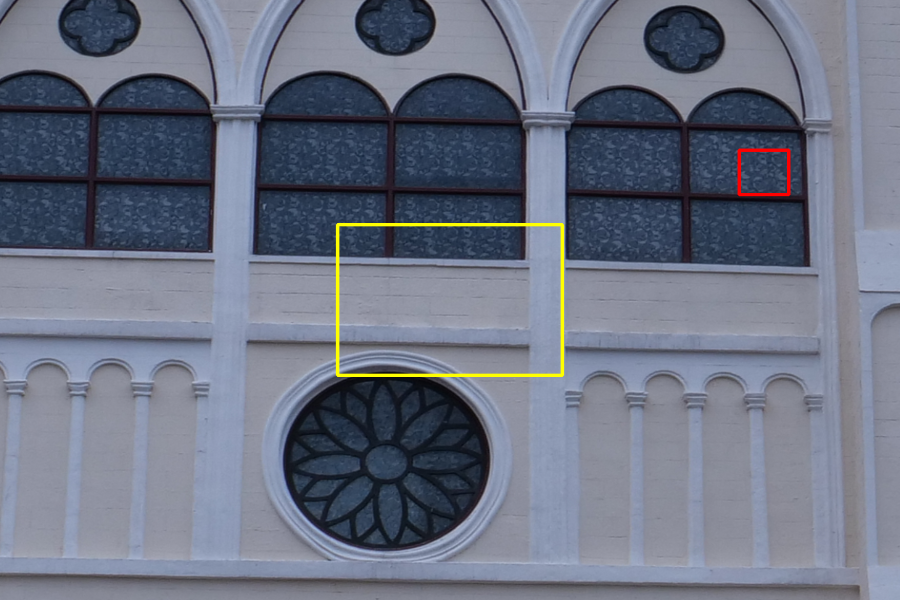}
        }
        \subfloat[][LR]
        {
            \includegraphics[width=.142\linewidth]{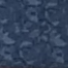}
        }
        \subfloat[][\cite{BSRGAN}]
        {
            \includegraphics[width=.142\linewidth]{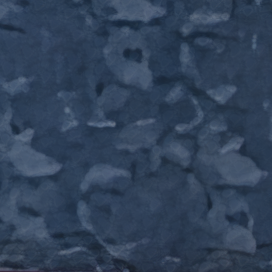}
        }
        \subfloat[][\cite{Real-ESRGAN}]
        {
            \includegraphics[width=.142\linewidth]{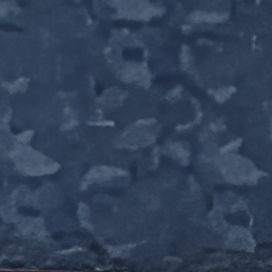}
        }
        \subfloat[][\cite{SRNTT}]
        {
            \includegraphics[width=.142\linewidth]{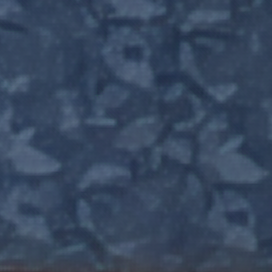}
        }
        \subfloat[][\cite{TTSR}]
        {
            \includegraphics[width=.142\linewidth]{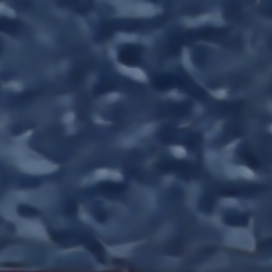}
        }
        
        \subfloat[][Telephoto]
        {
            \includegraphics[height=.142\linewidth, width=.213\linewidth]{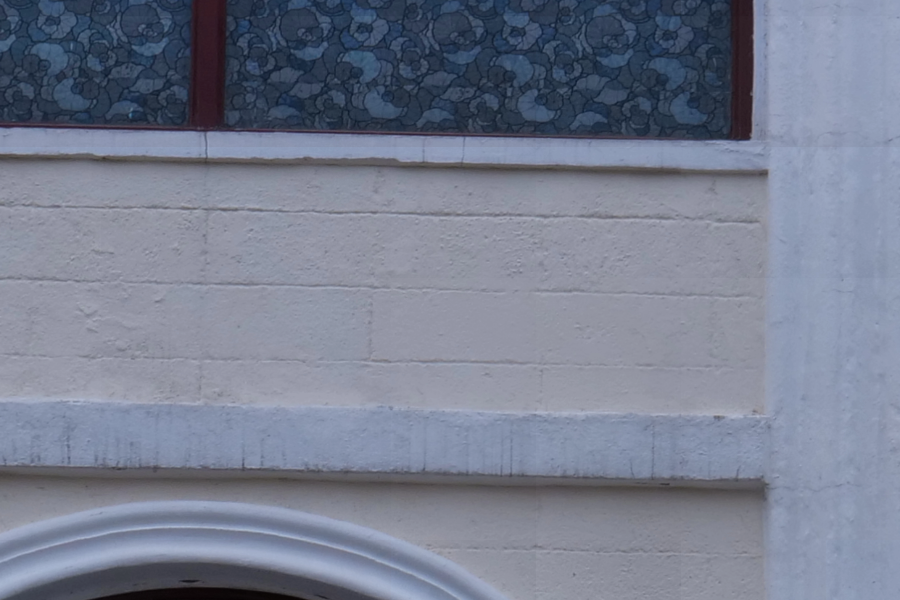}
        }
        \subfloat[][~\cite{C2-Matching}]
        {
            \includegraphics[width=.142\linewidth]{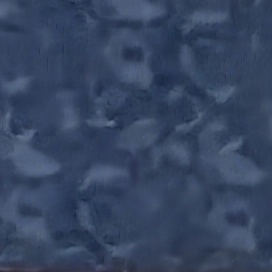}
        }
        \subfloat[][\cite{MASA-SR}]
        {
            \includegraphics[width=.142\linewidth]{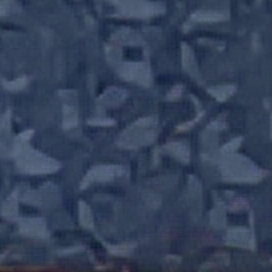}
        }
        \subfloat[][\cite{DCSR}]
        {
            \includegraphics[width=.142\linewidth]{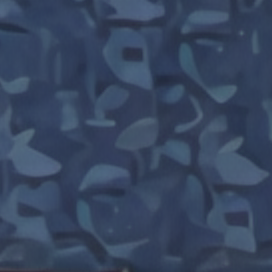}
        }
        \subfloat[][SelfDZSR]
        {
            \includegraphics[width=.142\linewidth]{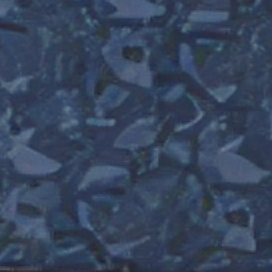}
        }
        \subfloat[][GT]
        {
            \includegraphics[width=.142\linewidth]{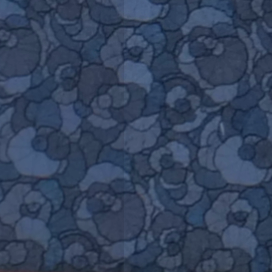}
        }
    \end{minipage}
    \caption{Visual comparison on \textbf{Panasonic} camera. In the short-focus image, the yellow box indicates the overlapped scene with the telephoto image, while the red box represents the selected LR patch. Our results in sub-figure (k) and (w) restore much more fine details.}
    \label{fig:Pan} 
\end{figure*}

\clearpage
%
%
\bibliographystyle{splncs04}
\bibliography{egbib}

\begin{thebibliography}{10}
\providecommand{\url}[1]{\texttt{#1}}
\providecommand{\urlprefix}{URL }
\providecommand{\doi}[1]{https://doi.org/#1}

\bibitem{KerGAN}
Bell-Kligler, S., Shocher, A., Irani, M.: Blind super-resolution kernel
  estimation using an internal-gan. In: NeurIPS. pp. 284--293 (2019)

\bibitem{NTIRE2019}
Cai, J., Gu, S., Timofte, R., Zhang, L.: Ntire 2019 challenge on real image
  super-resolution: Methods and results. In: CVPR Workshops. pp.~0--0 (2019)

\bibitem{LP-KPN}
Cai, J., Zeng, H., Yong, H., Cao, Z., Zhang, L.: Toward real-world single image
  super-resolution: A new benchmark and a new model. In: ICCV. pp. 3086--3095
  (2019)

\bibitem{chan2021understanding}
Chan, K.C., Wang, X., Yu, K., Dong, C., Loy, C.C.: Understanding deformable
  alignment in video super-resolution. In: AAAI. pp. 973--981 (2021)

\bibitem{CameraLen}
Chen, C., Xiong, Z., Tian, X., Zha, Z.J., Wu, F.: Camera lens super-resolution.
  In: CVPR. pp. 1652--1660 (2019)

\bibitem{DefConv}
Dai, J., Qi, H., Xiong, Y., Li, Y., Zhang, G., Hu, H., Wei, Y.: Deformable
  convolutional networks. In: ICCV. pp. 764--773 (2017)

\bibitem{SWLoss2}
Delbracio, M., Talebi, H., Milanfar, P.: Projected distribution loss for image
  enhancement. arXiv preprint arXiv:2012.09289  (2020)

\bibitem{SWDGen1}
Deshpande, I., Zhang, Z., Schwing, A.G.: Generative modeling using the sliced
  wasserstein distance. In: CVPR. pp. 3483--3491 (2018)

\bibitem{flow}
Dinh, L., Sohl-Dickstein, J., Bengio, S.: Density estimation using real nvp.
  In: ICLR (2017)

\bibitem{SRCNN}
Dong, C., Loy, C.C., He, K., Tang, X.: Image super-resolution using deep
  convolutional networks. IEEE PAMI  \textbf{38}(2),  295--307 (2015)

\bibitem{FlowNet}
Dosovitskiy, A., Fischer, P., Ilg, E., Hausser, P., Hazirbas, C., Golkov, V.,
  Van Der~Smagt, P., Cremers, D., Brox, T.: Flownet: Learning optical flow with
  convolutional networks. In: ICCV. pp. 2758--2766 (2015)

\bibitem{AdaConv}
Geng, Z., Sun, K., Xiao, B., Zhang, Z., Wang, J.: Bottom-up human pose
  estimation via disentangled keypoint regression. In: CVPR. pp. 14676--14686
  (2021)

\bibitem{GAN}
Goodfellow, I.J., Pouget-Abadie, J., Mirza, M., Xu, B., Warde-Farley, D.,
  Ozair, S., Courville, A.C., Bengio, Y.: Generative adversarial nets. In:
  NeurIPS (2014)

\bibitem{IKC}
Gu, J., Lu, H., Zuo, W., Dong, C.: Blind super-resolution with iterative kernel
  correction. In: CVPR. pp. 1604--1613 (2019)

\bibitem{he2020momentum}
He, K., Fan, H., Wu, Y., Xie, S., Girshick, R.: Momentum contrast for
  unsupervised visual representation learning. In: CVPR. pp. 9729--9738 (2020)

\bibitem{SWLoss1}
Heitz, E., Vanhoey, K., Chambon, T., Belcour, L.: A sliced wasserstein loss for
  neural texture synthesis. In: CVPR. pp. 9412--9420 (2021)

\bibitem{IMDN}
Hui, Z., Gao, X., Yang, Y., Wang, X.: Lightweight image super-resolution with
  information multi-distillation network. In: ACM MM. pp. 2024--2032 (2019)

\bibitem{correction_cvpr2020}
Hussein, S.A., Tirer, T., Giryes, R.: Correction filter for single image
  super-resolution: Robustifying off-the-shelf deep super-resolvers. In: CVPR.
  pp. 1428--1437 (2020)

\bibitem{STN}
Jaderberg, M., Simonyan, K., Zisserman, A., et~al.: Spatial transformer
  networks. In: NeurIPS. pp. 2017--2025 (2015)

\bibitem{C2-Matching}
Jiang, Y., Chan, K.C., Wang, X., Loy, C.C., Liu, Z.: Robust reference-based
  super-resolution via c2-matching. In: CVPR. pp. 2103--2112 (2021)

\bibitem{RelateGAN}
Jolicoeur-Martineau, A.: The relativistic discriminator: a key element missing
  from standard gan. arXiv preprint arXiv:1807.00734  (2018)

\bibitem{Adam}
Kingma, D.P., Ba, J.: Adam: A method for stochastic optimization. In: ICLR
  (2015)

\bibitem{ClassSR}
Kong, X., Zhao, H., Qiao, Y., Dong, C.: Classsr: A general framework to
  accelerate super-resolution networks by data characteristic. In: CVPR. pp.
  12016--12025 (2021)

\bibitem{SRGAN}
Ledig, C., Theis, L., Husz{\'a}r, F., Caballero, J., Cunningham, A., Acosta,
  A., Aitken, A., et~al.: Photo-realistic single image super-resolution using a
  generative adversarial network. In: CVPR. pp. 4681--4690 (2017)

\bibitem{MANet}
Liang, J., Sun, G., Zhang, K., Van~Gool, L., Timofte, R.: Mutual affine network
  for spatially variant kernel estimation in blind image super-resolution. In:
  ICCV. pp. 4096--4105 (2021)

\bibitem{FKP}
Liang, J., Zhang, K., Gu, S., Van~Gool, L., Timofte, R.: Flow-based kernel
  prior with application to blind super-resolution. In: CVPR. pp. 10601--10610
  (2021)

\bibitem{EDSR}
Lim, B., Son, S., Kim, H., Nah, S., Mu~Lee, K.: Enhanced deep residual networks
  for single image super-resolution. In: CVPR Workshops. pp. 136--144 (2017)

\bibitem{AdaDSR}
Liu, M., Zhang, Z., Hou, L., Zuo, W., Zhang, L.: Deep adaptive inference
  networks for single image super-resolution. In: ECCV Workshops. pp. 131--148.
  Springer (2020)

\bibitem{MASA-SR}
Lu, L., Li, W., Tao, X., Lu, J., Jia, J.: Masa-sr: Matching acceleration and
  spatial adaptation for reference-based image super-resolution. In: CVPR. pp.
  6368--6377 (2021)

\bibitem{NTIRE2020}
Lugmayr, A., Danelljan, M., Timofte, R.: Ntire 2020 challenge on real-world
  image super-resolution: Methods and results. In: CVPR Workshops. pp. 494--495
  (2020)

\bibitem{AIM2019}
Lugmayr, A., Danelljan, M., Timofte, R., Fritsche, M., et~al.: Aim 2019
  challenge on real-world image super-resolution: Methods and results. In: ICCV
  Workshops. pp. 3575--3583. IEEE (2019)

\bibitem{DAN}
Luo, Z., Huang, Y., Li, S., Wang, L., Tan, T.: Unfolding the alternating
  optimization for blind super resolution. In: NeurIPS (2020)

\bibitem{PyTorch}
Paszke, A., Gross, S., Massa, F., Lerer, A., Bradbury, J., Chanan, G., Killeen,
  T., Lin, Z., Gimelshein, N., Antiga, L., Desmaison, A., Kopf, A., Yang, E.,
  DeVito, Z., Raison, M., Tejani, A., Chilamkurthy, S., Steiner, B., Fang, L.,
  Bai, J., Chintala, S.: Pytorch: An imperative style, high-performance deep
  learning library. In: NeurIPS. pp. 8024--8035 (2019)

\bibitem{PixelShuffle}
Shi, W., Caballero, J., Husz{\'a}r, F., Totz, J., Aitken, A.P., Bishop, R.,
  Rueckert, D., Wang, Z.: Real-time single image and video super-resolution
  using an efficient sub-pixel convolutional neural network. In: CVPR. pp.
  1874--1883 (2016)

\bibitem{SSEN}
Shim, G., Park, J., Kweon, I.S.: Robust reference-based super-resolution with
  similarity-aware deformable convolution. In: CVPR. pp. 8425--8434 (2020)

\bibitem{VGGLOSS}
Simonyan, K., Zisserman, A.: Very deep convolutional networks for large-scale
  image recognition. In: ICLR (2014)

\bibitem{PWC-Net}
Sun, D., Yang, X., Liu, M.Y., Kautz, J.: Pwc-net: Cnns for optical flow using
  pyramid, warping, and cost volume. In: CVPR. pp. 8934--8943 (2018)

\bibitem{wang2021exploring}
Wang, L., Dong, X., Wang, Y., Ying, X., Lin, Z., An, W., Guo, Y.: Exploring
  sparsity in image super-resolution for efficient inference. In: CVPR. pp.
  4917--4926 (2021)

\bibitem{DASR}
Wang, L., Wang, Y., Dong, X., Xu, Q., Yang, J., An, W., Guo, Y.: Unsupervised
  degradation representation learning for blind super-resolution. In: CVPR. pp.
  10581--10590 (2021)

\bibitem{DCSR}
Wang, T., Xie, J., Sun, W., Yan, Q., Chen, Q.: Dual-camera super-resolution
  with aligned attention modules. In: ICCV. pp. 2001--2010 (2021)

\bibitem{Real-ESRGAN}
Wang, X., Xie, L., Dong, C., Shan, Y.: Real-esrgan: Training real-world blind
  super-resolution with pure synthetic data. In: ICCV Workshops. pp. 1905--1914
  (2021)

\bibitem{SSIM}
Wang, Z., Bovik, A.C., Sheikh, H.R., Simoncelli, E.P.: Image quality
  assessment: from error visibility to structural similarity. IEEE TIP
  \textbf{13}(4),  600--612 (2004)

\bibitem{AIM2020}
Wei, P., Lu, H., Timofte, R., Lin, L., Zuo, W., et~al.: Aim 2020 challenge on
  real image super-resolution: methods and results. In: ECCV Workshops. pp.
  392--422. Springer (2020)

\bibitem{CDC}
Wei, P., Xie, Z., Lu, H., Zhan, Z., Ye, Q., Zuo, W., Lin, L.: Component
  divide-and-conquer for real-world image super-resolution. In: ECCV. pp.
  101--117. Springer (2020)

\bibitem{wei2021unsupervised}
Wei, Y., Gu, S., Li, Y., Timofte, R., Jin, L., Song, H.: Unsupervised
  real-world image super resolution via domain-distance aware training. In:
  CVPR. pp. 13385--13394 (2021)

\bibitem{SWDGen2}
Wu, J., Huang, Z., Acharya, D., Li, W., Thoma, J., Paudel, D.P., Gool, L.V.:
  Sliced wasserstein generative models. In: CVPR. pp. 3713--3722 (2019)

\bibitem{Xie_2021_ICCV}
Xie, W., Song, D., Xu, C., Xu, C., Zhang, H., Wang, Y.: Learning
  frequency-aware dynamic network for efficient super-resolution. In: ICCV. pp.
  4308--4317 (2021)

\bibitem{FRM}
Xie, Y., Xiao, J., Sun, M., Yao, C., Huang, K.: Feature representation matters:
  End-to-end learning for reference-based image super-resolution. In: ECCV. pp.
  230--245. Springer (2020)

\bibitem{TTSR}
Yang, F., Yang, H., Fu, J., Lu, H., Guo, B.: Learning texture transformer
  network for image super-resolution. In: CVPR. pp. 5791--5800 (2020)

\bibitem{BSRGAN}
Zhang, K., Liang, J., Van~Gool, L., Timofte, R.: Designing a practical
  degradation model for deep blind image super-resolution. In: ICCV. pp.
  4791--4800 (2021)

\bibitem{SRMD}
Zhang, K., Zuo, W., Zhang, L.: Learning a single convolutional super-resolution
  network for multiple degradations. In: CVPR. pp. 3262--3271 (2018)

\bibitem{LPIPS}
Zhang, R., Isola, P., Efros, A.A., Shechtman, E., Wang, O.: The unreasonable
  effectiveness of deep features as a perceptual metric. In: CVPR. pp. 586--595
  (2018)

\bibitem{SRRAW}
Zhang, X., Chen, Q., Ng, R., Koltun, V.: Zoom to learn, learn to zoom. In:
  CVPR. pp. 3762--3770 (2019)

\bibitem{RCAN}
Zhang, Y., Li, K., Li, K., Wang, L., Zhong, B., Fu, Y.: Image super-resolution
  using very deep residual channel attention networks. In: ECCV. pp. 286--301
  (2018)

\bibitem{zhang2020texture}
Zhang, Y., Zhang, Z., DiVerdi, S., Wang, Z., Echevarria, J., Fu, Y.: Texture
  hallucination for large-factor painting super-resolution. In: ECCV. pp.
  209--225 (2020)

\bibitem{SRNTT}
Zhang, Z., Wang, Z., Lin, Z., Qi, H.: Image super-resolution by neural texture
  transfer. In: CVPR. pp. 7982--7991 (2019)

\bibitem{RAW-to-sRGB}
Zhang, Z., Wang, H., Liu, M., Wang, R., Zhang, J., Zuo, W.: Learning
  raw-to-srgb mappings with inaccurately aligned supervision. In: ICCV. pp.
  4348--4358 (2021)

\bibitem{refsr_bwvc}
Zheng, H., Ji, M., Han, L., Xu, Z., Wang, H., Liu, Y., Fang, L.: Learning
  cross-scale correspondence and patch-based synthesis for reference-based
  super-resolution. In: BMVC. vol.~1, p.~2 (2017)

\bibitem{CrossNet}
Zheng, H., Ji, M., Wang, H., Liu, Y., Fang, L.: Crossnet: An end-to-end
  reference-based super resolution network using cross-scale warping. In: ECCV.
  pp. 88--104 (2018)

\bibitem{DefConv-v2}
Zhu, X., Hu, H., Lin, S., Dai, J.: Deformable convnets v2: More deformable,
  better results. In: CVPR. pp. 9308--9316 (2019)

\end{thebibliography}


\end{document}